\documentclass[twocolumn,nofootinbib,amsmath,amssymb]{revtex4}
\usepackage{graphicx}

\newcommand{\be}{\begin{equation}}
\newcommand{\ee}{\end{equation}}
\newcommand{\ba}{\begin{eqnarray}}
\newcommand{\ea}{\end{eqnarray}}
\newcommand{\hs}{\hspace{-1.12mm}}
\newcommand{\eq}{\hs & = & \hs}
\newcommand{\eqequiv}{\hs & \equiv & \hs}
\newcommand{\eqsimeq}{\hs & \simeq & \hs}
\newcommand{\eqnl}{\hs & & \hs}
\newcommand{\vs}{\vspace{1.5mm}}
\newcommand{\addspace}[2][3mm]{\raisebox{0mm}[#1][#1]{$\displaystyle #2$}}
\newcommand{\addsmsp}[1]{\addspace[2mm]{#1}}
\newcommand{\lcdm}{\Lambda\mathrm{CDM}}
\newcommand{\const}{\mathrm{const}}
\newcommand{\kelvin}{\mathrm{K}}
\newcommand{\meter}{\mathrm{m}}
\newcommand{\second}{\mathrm{s}}
\newcommand{\hydr}{\mathrm{H}}
\newcommand{\he}{\mathrm{He}}
\newcommand{\ev}{\mathrm{eV}}

\begin{document}

\title{How to calculate the CMB spectrum}

\author{Petter Callin}
  \email{n.p.callin@fys.uio.no}
\affiliation{Department of Physics, University of Oslo, N-0316 Oslo, Norway\\}

\date{June 28, 2006}

\begin{abstract}
We present a self-contained description of everything needed to write a program
that calculates the CMB power spectrum for the standard model of cosmology
($\lcdm$). This includes the equations used, assumptions and approximations
imposed on their solutions, and most importantly the algorithms and programming
tricks needed to make the code actually work. The resulting program is compared
to CMBFAST and typically agrees to within $0.1 \,\%$ -- $0.4 \,\%$. It includes
both helium, reionization, neutrinos and the polarization power spectrum. The
methods presented here could serve as a starting point for people wanting to
write their own CMB program from scratch, for instance to look at more exotic
cosmological models where CMBFAST or the other standard programs can't be used
directly.
\end{abstract}

\maketitle

\section{Introduction}

Since the first detection of anisotropies in the Cosmic Microwave Background
(CMB) by the Cosmic Background Explorer (COBE) satellite in the early 90's
\cite{COBE}, there has been considerable activity in this field of cosmology
throughout the world. With the more accurate measurements of the Wilkinson
Microwave Anisotropy Probe (WMAP) \cite{WMAP_firstyear, WMAP_threeyear} and the
future Planck satellite \cite{Planck}, we are entering the era of precision
cosmology. Results from different kinds of experiments seem to converge to what
is being referred to as the Standard Model of Cosmology \cite{StandardModel},
describing the history of the universe from inflation through Big Bang
Nucleosynthesis to the release of the CMB radiation during recombination. With
the growing precision of the observational data comes also the need for fast
and accurate theoretical calculations of CMB power spectra. Often one seeks the
best fit to observations of a model with several parameters, requiring
typically hundreds of spectra to be calculated and compared to the data.

Of course, several standard computer programs that calculate CMB power spectra
are already available. The most commonly used include CMBFAST \cite{CMBFAST,
CMBFAST_web}, CMBEASY\- \cite{CMBEASY}, and CAMB \cite{CAMB}. These are all
excellent programs when calculating power spectra for the $\lcdm$ model,
including also some extensions like a quintessence field, hot dark matter or a
simple fifth dimension. One may therefore wonder what the point is writing a
new program from scratch, a task which obviously requires quite a lot of work.
The need may arise when considering more exotic cosmological models which are
not included in any of the standard programs. This could for instance include
quintessence with a non-trivial coupling to other fields, extra dimensions with
non-trivial geometry, or a model with varying constants of nature. In these
cases one has the choice of either extending existing code, or writing a
completely new program. The former can certainly be the easiest solution in
some situations, but not necessarily all. The problem with updating existing
code is that one probably doesn't know exactly how it works, and it is
therefore difficult to make changes without doing something wrong. By instead
writing new code, starting with a simple model and comparing the results to
existing programs, it will be much easier to later extend the program since one
then knows precisely what every line of code is doing. By knowing the code in
detail, more confidence can also be put in the results. Obviously a lot of work
is needed to write a CMB program from scratch. Fortunately this work is not a
complete waste of time, since a lot of new insight about the underlying physics
can be obtained in the process.

The purpose of this text is to provide a self-contained collection of all the
ingredients needed to write a program that calculates the CMB power spectrum.
This includes both the equations governing the physics, any assumptions or
approximations used in their solutions, and pointing out what algorithms and
tricks to use when implementing the equations in a computer program. The main
focus will be the practical computer implementation of the equations, not their
derivations, for which we instead point the reader to the references. Only new
or less readily available derivations will be included. We are only assuming
that the reader has some basic knowledge of CMB physics, and some experience
with a high-level programming language~\footnote{We will not show any code
written in a spesific programming language, only the general algorithms used.
For the record, we have used Delphi for Windows \cite{Delphi} in this work, but
C or Fortran (or similar languages) are equally well suited.}. Most of the
presentation will follow the notation and conventions of
Dodelson~\cite{Dodelson}.

In section \ref{cha:background} we start by going through the background
cosmology, which includes background geometry and the recombination history of
the universe, and in section \ref{cha:perturbations} we introduce perturbations
to this background. The equations of motion for the perturbations are given by
the various Boltzmann equations. We then state the initial conditions used, and
the approximations used during tight coupling. In section \ref{cha:cmbspectrum}
we go from the perturbations of the CMB temperature to the spectrum of $C_l$'s.
The most important programming techniques are mentioned in section
\ref{cha:programming}, including the cutoff scheme for the Boltzmann
hierarchies, how to integrate the various equations numerically, and the
normalization of the spectrum. The resulting spectra are compared to CMBFAST in
section \ref{cha:results}. In section \ref{cha:more_ingredients} the program is
extended to include a few more effects, including helium, a simple model of
reionization, massless neutrinos, and the (E-mode) polarization power spectrum.
Finally we conclude in section~\ref{cha:conclusion}.

\section{Background cosmology}
\label{cha:background}

Before looking at perturbations, we must determine the background. This
consists of two parts: The easiest part is the background geometry, which is
given by the standard Friedmann-Robertson-Walker (FRW) metric. The more
difficult part is the recombination history of the universe, which involves
finding the number of free electrons and electrons bound to neutral atoms as a
function of time. We need this to determine the coupling between photons and
baryons.

\subsection{Background geometry}

The background geometry is given by the FRW metric
\ba
  ds^2 \eq -dt^2 + a^2(t) \delta_{ij} dx^i dx^j \nonumber \\
  \eq a^2(\eta) \left( -d\eta^2 + \delta_{ij} dx^i dx^j \right) ,
\ea
where $t$ is the physical time and $\eta$ the conformal time, and $a$ is the
scale factor describing the expansion of the universe, which we assume is
spatially flat ($k=0$). The expansion is given by Friedmann's equation
\ba
  H \eqequiv \frac{1}{a} \frac{da}{dt} =
    H_0 \sqrt{
      (\Omega_m + \Omega_b) a^{-3} + \Omega_r a^{-4} + \Omega_\Lambda
    } \, , \nonumber \\
  {\cal H} \eqequiv \frac{1}{a} \frac{da}{d\eta} \equiv
    \frac{\dot{a}}{a} = a H \nonumber \\
  \eq H_0 \sqrt{
      (\Omega_m + \Omega_b) a^{-1} + \Omega_r a^{-2} + \Omega_\Lambda a^2
    } \, ,
  \label{eq:Hubble}
\ea
where the dot means the derivative with respect to conformal time, and we
assume that the universe consists of cold dark matter (CDM, $m$), baryons
$(b)$, radiation $(r)$, and a cosmological constant $(\Lambda)$. $H_0$ is the
current value of the Hubble constant. We also introduce the logarithm of the
scale factor,
\be
  x \equiv \ln a \, , \qquad ' \equiv \frac{d}{dx} \, .
\ee
The various "time variables" $t$, $\eta$, $a$ and $x$ are related through the
useful equations
\ba
  \frac{dt}{d\eta} = a \, , \quad \frac{dx}{dt} = H \, , \quad
  \frac{dx}{d\eta} = {\cal H} \, , && \nonumber \\
  \frac{d}{dt} = H \frac{d}{dx} \, , \quad
  \frac{d}{d\eta} = {\cal H} \frac{d}{dx} \, . &&
\ea
We will also need an expression for the conformal time as a function of the
scale factor in our calculations:
\be
  \eta(a) = \int_0^a \frac{da'}{a' {\cal H}(a')} \, , \qquad
  \eta(x) = \eta(a = e^x) \, .
  \label{eq:eta_a}
\ee
This integral is easily calculated numerically. Note that $a {\cal H}(a) \to
H_0 \sqrt{\Omega_r}$ as $a \to 0$, so there's no problem with convergence.

\subsection{Recombination}
\label{cha:recombination}

In the early universe all atoms were fully ionized, giving a strong coupling
between the baryon and photon plasma due to Thomson scattering. When the
temperature dropped below $\sim 3000 \,\kelvin$ neutral atoms were formed, and
the universe became transparent. The CMB photons we observe today have
travelled more or less freely through the universe since they were last
scattered during recombination. The optical depth $\tau$ back to conformal time
$\eta$ is given by~\cite{Dodelson}
\ba
  \tau(\eta) \eq \int_\eta^{\eta_0} n_e \sigma_T a \, d\eta' \, , \nonumber \\
  \dot{\tau} = -n_e \sigma_T a \, \quad &\Leftrightarrow& \quad
  \tau' = -\frac{n_e \sigma_T a}{{\cal H}} \, ,
\ea
where $n_e$ is the number density of free electrons,
\be
  \sigma_T = \frac{8\pi\alpha^2}{3 m_e^2} =
    6.652462 \times 10^{-29} \; \meter^2
\ee
the Thomson cross section, and $\eta_0$ the conformal time today, $\eta_0 =
\eta(a=1)$. We define the visibility function
\ba
  g(\eta) \eq -\dot{\tau} e^{-\tau(\eta)} =
    -{\cal H} \tau' e^{-\tau(x)} = g(x) \, , \nonumber \\
  \tilde{g}(x) \eqequiv -\tau' e^{-\tau} = \frac{g(x)}{{\cal H}(x)} \, .
\ea
The visibility function is normalized as
\be
  \int_0^{\eta_0} g(\eta) d\eta =
    \int_{-\infty}^0 \tilde{g}(x) dx = 1 \, ,
\ee
and can therefore be interpreted as a probability distribution, namely the
probability that a CMB photon observed today was last scattered at conformal
time $\eta$. The function $g$ has a relatively sharp peak at a certain
redshift, of order $z \sim 1100$, which we therefore call the time of
recombination. Most CMB photons were last scattered around this time.

The difficult task is to calculate the electron density $n_e$. We define the
free electron fraction
\be
  X_e \equiv \frac{n_e}{n_\hydr} = \frac{n_e}{n_b} \, ,
  \label{eq:electronfrac}
\ee
where the total number density of hydrogen, $n_\hydr$, is equal to the baryon
number density $n_b$ when we ignore helium (see section \ref{cha:helium}).
Ignoring also the small mass difference between free protons and neutral
hydrogen, we have
\be
  n_\hydr = n_b \simeq \frac{\rho_b}{m_\hydr} =
    \frac{\Omega_b \rho_c}{m_\hydr a^3} \, ,
  \qquad \rho_c \equiv \frac{3 H_0^2}{8\pi G} \, .
\ee
Here $m_\hydr$ is the mass of the hydrogen atom, and $\rho_c$ the critical
density today. At early times all hydrogen is completely ionized, so $X_e
\simeq 1$ and $n_e \sim a^{-3}$, whereas at late times $X_e \ll 1$ (but does
not approach zero).

Before recombination the electron fraction can be approximated by the Saha
equation~\cite{Dodelson, Ma_Bert, Hu}:
\be
  \frac{X_e^2}{1-X_e} \simeq \frac{1}{n_b}
    \left( \frac{m_e T_b}{2\pi} \right)^{3/2} e^{-\epsilon_0/T_b} \, .
  \label{eq:Saha}
\ee
Here $T_b$ is the baryon temperature, and $\epsilon_0 = 13.605698 \; \ev$ the
ionization energy of hydrogen. During and after recombination, however, we must
use the more accurate Peebles equation~\cite{Dodelson, Ma_Bert}
\be
  \frac{dX_e}{dx} = \frac{C_r(T_b)}{H} \left[
    \beta(T_b) (1-X_e) - n_\hydr \alpha^{(2)}(T_b) X_e^2
  \right] ,
  \label{eq:Peebles}
\ee
where
\ba
  C_r(T_b) \eq \frac{\Lambda_{2s \to 1s} + \Lambda_\alpha}
    {\Lambda_{2s \to 1s} + \Lambda_\alpha + \beta^{(2)}(T_b)} \, , \nonumber \\
  \Lambda_{2s \to 1s} \eq 8.227 \; \second^{-1} \, ,
    \qquad \Lambda_\alpha = H \frac{(3\epsilon_0)^3}{(8\pi)^2 n_{1s}} \, ,
    \nonumber \\
  n_{1s} \eqsimeq (1-X_e) n_\hydr \, ,
    \qquad \beta^{(2)}(T_b) = \beta(T_b) e^{3\epsilon_0/4T_b} \, , \nonumber \\
  \beta(T_b) \eq \alpha^{(2)}(T_b) \left( \frac{m_e T_b}{2\pi} \right)^{3/2}
    e^{-\epsilon_0/T_b} \, , \nonumber \\
  \alpha^{(2)}(T_b) \eq \frac{64\pi}{\sqrt{27\pi}} \frac{\alpha^2}{m_e^2}
    \sqrt{\frac{\epsilon_0}{T_b}} \phi_2(T_b) \, , \nonumber \\
  \phi_2(T_b) \eqsimeq 0.448 \ln(\epsilon_0/T_b) \, .
  \label{eq:Peebles_def}
\ea
The various terms here are described in more detail in \cite{Ma_Bert}. The
baryon temperature $T_b$ has a non-trivial time evolution, and is given by a
differential equation which couples to $X_e$ \cite{Ma_Bert}. Thus, we actually
have a complicated coupled system of differential equations for both $X_e$ and
$T_b$. However, the error when setting the baryon temperature equal to the
photon temperature throughout recombination turns out to be only of order
$10^{-6}$ \cite{Reijo}. We therefore use the approximation
\be
  T_b \simeq T_r = \frac{T_0}{a} \, , \qquad
  T_0 = 2.725 \; \kelvin \, .
\ee
Naively, one would probably think that recombination occurs when $T \sim
\epsilon_0 \simeq 157\,900 \;\kelvin$ when looking at (\ref{eq:Saha}) or
(\ref{eq:Peebles}). However, it is delayed until $T \simeq 3000 \;\kelvin$
because of the large photon to baryon number ratio.

\begin{figure}[!h]
  \begin{center}
    \includegraphics{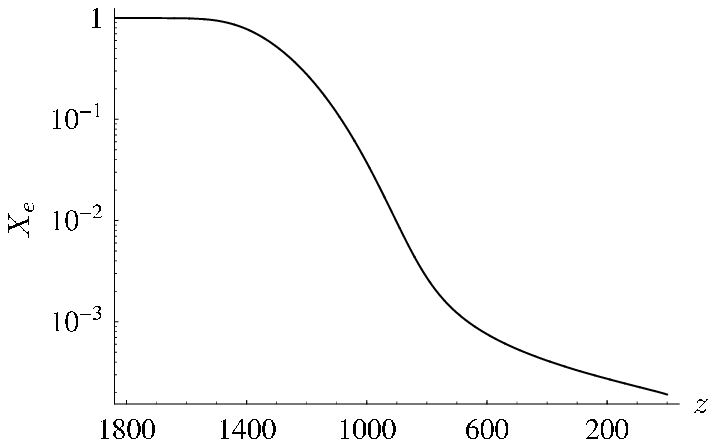}
  \end{center}
  \vspace{-6mm}
  \caption{The free electron fraction $X_e$ as a function of redshift, using
  the Saha approximation (\ref{eq:Saha}) until $z =  1587.4$ where $X_e =
  0.99$, and then integrating the Peebles equation~(\ref{eq:Peebles}).}
  \label{fig:electronfrac}
\end{figure}

\begin{figure}[!h]
  \begin{center}
    \includegraphics{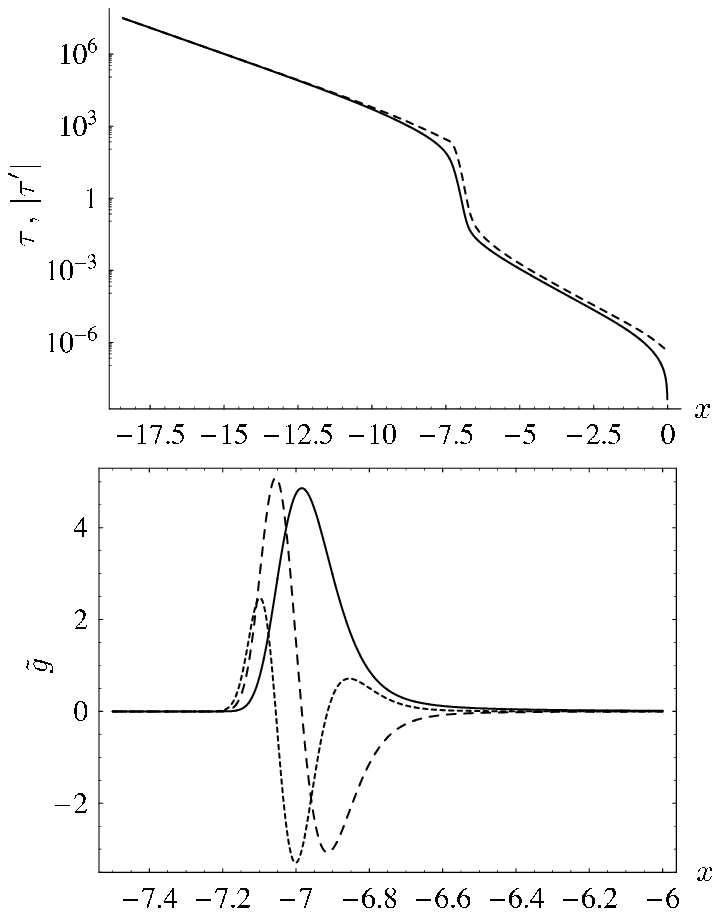}
  \end{center}
  \vspace{-6mm}
  \caption{The top figure shows the optical depth $\tau$ (solid line)
  and $|\tau'|$ (dashed line) as functions of $x$. The bottom figure
  shows the visibility function $\tilde{g} = -\tau' e^{-\tau}$
  (solid line), its derivative $\tilde{g}'/10$ (dashed line), and its double
  derivative $\tilde{g}''/300$ (dotted line), also as functions of $x$. The
  scaling is chosen to make the curves fit into the same figure. The visibility
  function has a peak at $x = -6.984$, corresponding to a redshift $z = 1078$.
  The peak of $\tilde{g}$ has both a finite width and a clear asymmetry, and is
  therefore not particularly well approximated by a delta function.}
  \label{fig:tau_g}
\end{figure}

It is difficult to integrate the Peebles equation (\ref{eq:Peebles})
numerically at very early times, but this is also the place where the Saha
equation (\ref{eq:Saha}) is a good approximation. In the numerical calculation
we therefore use Saha until the electron fraction $X_e$ has been reduced to,
say, $0.99$, and then switch to Peebles using Saha as the initial condition.
Figure \ref{fig:electronfrac} shows the numerical solution $X_e$ as a function
of redshift $z = a^{-1} - 1$, and figure \ref{fig:tau_g} the resulting optical
depth and visibility function, all for the model
\be
  \begin{array}{rrcl}
    \mbox{Hubble constant:} & h \eq 0.7 \, , \vs \\
      & \hspace{-30mm} (H_0 \eq h \cdot 100 \;
        \mathrm{km \, s^{-1} \, Mpc^{-1}}) \hspace{-2mm} \vs \\
    \mbox{CMB temperature:} & T_0 \eq 2.725 \; \kelvin \, , \vs \\
      & \hspace{-30mm} (\, \Rightarrow \; \Omega_r \eq
        5.042 \times 10^{-5}) \vs \\
    \mbox{Baryon density:} & \Omega_b \eq 0.046 \, , \vs \\
    \mbox{CDM density:} & \Omega_m \eq 0.224 \, , \vs \\
    \mbox{Spectral index:} & n \eq 1 \, , \vs \\
    \mbox{Helium mass fraction:} & Y_p \eq 0 \, , \vs \\
    \mbox{Vacuum density:} & \Omega_\Lambda \eq
      1 - (\Omega_r + \Omega_b + \Omega_m) \vs \\
    & \eq 0.72995 \, ,
  \end{array} \hspace{-5mm}
  \label{eq:defaultmodel}
\ee
which we call the "default model" for the rest of this text.

\section{Perturbations}
\label{cha:perturbations}

\subsection{Definitions}

Having determined the background cosmology, we can now turn to the
perturbations. We use the Newtonian gauge and write the perturbed metric~as
\be
  g_{\mu\nu} = \left( \begin{array}{cc}
    -(1 + 2\Psi) & 0 \vs \\
    0 & a^2 \delta_{ij} (1 + 2\Phi)
  \end{array} \right) .
\ee
We are therefore only considering scalar perturbations. Perturbations to the
photons are defined as the relative variation of the photon temperature:
\be
  T(\vec{k}, \mu, \eta) = T^{(0)}(\eta) \left[
    1 + \Theta(\vec{k}, \mu, \eta)
  \right] , \quad \mu = \frac{\vec{k}\cdot\vec{p}}{kp} \, .
\ee
We will be working in Fourier space throughout this text, with $\vec{k}$ as the
Fourier transformed variable of the position $\vec{x}$. The momentum of the
photon itself is $\vec{p}$. Note that the perturbation $\Theta$ depends only on
the direction of $\vec{p}$, not its magnitude. The direction dependence is what
leads to anisotropies in the CMB field. The photon perturbation is expanded in
multipoles:
\ba
  \Theta_l \eq \frac{i^l}{2} \int_{-1}^1 {\cal P}_l(\mu) \Theta(\mu) d\mu \, ,
    \nonumber \\
  \Leftrightarrow \qquad \Theta(\mu) \eq
    \sum_{l=0}^\infty \frac{2l+1}{i^l} \Theta_l {\cal P}_l(\mu) \, ,
\ea
where ${\cal P}_l(\mu)$ are the Legendre polynomials. In addition to the
temperature perturbation, there's also perturbations to the photon
polarization, which we denote by $\Theta_P(\mu)$. (See section
\ref{cha:polarization} for more details on polarization.)

\subsection{Perturbation equations}

The equations of motion for the perturbations follow from the Boltzmann
equations for photons, CDM, baryons and neutrinos. Since CDM and baryons are
non-relativistic, we can take the first two moments of their equations instead
of keeping an arbitrary direction dependence, obtaining equations for the
density and velocity. In addition to the Boltzmann equations, Einsteins
equation gives two equations for the two gravitational potentials. In total,
the evolution of the perturbations is governed by the following system of
equations~\cite{Dodelson}:
\ba
  \dot{\Theta} + ik\mu\Theta \eq -\dot{\Phi} - ik\mu\Psi \nonumber \\
  \eqnl - \, \dot{\tau} \left[
    \Theta_0 - \Theta + i\mu v_b - \frac{1}{2} {\cal P}_2 \Pi
  \right] , \nonumber \\
  \dot{\Theta}_P + ik\mu\Theta_P \eq -\dot{\tau} \left[
    -\Theta_P + \frac{1}{2} (1 - {\cal P}_2) \Pi
  \right] , \nonumber \\
  \dot{\delta} - kv \eq -3\dot{\Phi} \, , \qquad
  \dot{v} + {\cal H} v = -k\Psi \, , \nonumber \\
  \dot{\delta}_b - kv_b \eq -3\dot{\Phi} \, , \nonumber \\
  \dot{v}_b + {\cal H} v_b \eq -k\Psi +
    \dot{\tau} R \left( v_b + 3\Theta_1 \right) , \nonumber \\
  \dot{\cal N} + ik\mu {\cal N} \eq -\dot{\Phi} - ik\mu\Psi \, , \nonumber \\
  k^2 \Phi + 3 {\cal H} \left( \dot{\Phi} - {\cal H} \Psi \right) \eq
    4\pi G a^2 \left[ \addsmsp{ \rho\delta + \rho_b \delta_b } \right.
    \nonumber \\
  && \hspace{10.6mm} \left. \addsmsp{
    + \, 4\rho_r \Theta_0 + 4\rho_\nu {\cal N}_0
    } \right] , \nonumber \\
  k^2 (\Phi+\Psi) \eq -32\pi G a^2
    \left[ \addsmsp{ \rho_r \Theta_2 + \rho_\nu {\cal N}_2 } \right] ,
    \nonumber \\
  && \hspace{-35mm} \Pi = \Theta_2 + \Theta^P_2 + \Theta^P_0 \, , \qquad
  R = \frac{4\rho_r}{3\rho_b} = \frac{4\Omega_r}{3\Omega_b a} \, .
  \label{eq:diffeqs_start}
\ea
Here $\delta$ and $v$ are the density perturbation and velocity of CDM,
$\delta_b$ and $v_b$ the same for baryons, and ${\cal N}$ (massless) neutrino
perturbations. Compared to \cite[eqs. 4.100 -- 4.107]{Dodelson} we have defined
$R \to 1/R$, and $v \to iv$, $v_b \to i v_b$ to make the velocities real.
Expanding in multipoles, the equations for $\Theta$, $\Theta_P$ and ${\cal N}$
turn into the hierarchies
\ba
  \dot{\Theta}_0 + k\Theta_1 \eq -\dot{\Phi} \, , \nonumber \\
  \dot{\Theta}_1 - \frac{k}{3} \Theta_0 + \frac{2k}{3} \Theta_2 \eq
    \frac{k}{3} \Psi + \dot{\tau}
    \!\left[ \Theta_1 \!+\! \frac{1}{3} v_b \right] \! ,
    \nonumber \\
  \dot{\Theta}_l - \frac{lk}{2l+1} \Theta_{l-1} +
    \frac{(l+1)k}{2l+1} \Theta_{l+1} \eq \dot{\tau}
    \left[ \Theta_l - \frac{1}{10} \Pi \, \delta_{l,2} \right] \! ,
    \nonumber \\
  \eqnl (l \geq 2) \nonumber \\
  \dot{\Theta}^P_0 + k \Theta^P_1 \eq
    \dot{\tau} \left[ \Theta^P_0 - \frac{1}{2} \Pi \right] \! , \nonumber \\
  \dot{\Theta}^P_l - \frac{lk}{2l+1} \Theta^P_{l-1} +
    \frac{(l+1)k}{2l+1} \Theta^P_{l+1} \eq \dot{\tau}
      \left[ \Theta^P_l - \frac{1}{10} \Pi \, \delta_{l,2} \right] \! ,
    \nonumber \\
  \eqnl (l \geq 1) \nonumber \\
  \dot{\cal N}_0 + k {\cal N}_1 \eq -\dot{\Phi} \, , \nonumber \\
  \dot{\cal N}_1 - \frac{k}{3} {\cal N}_0 + \frac{2k}{3} {\cal N}_2 \eq
    \frac{k}{3} \Psi \, , \nonumber \\
  \dot{\cal N}_l - \frac{lk}{2l+1} {\cal N}_{l-1} +
    \frac{(l+1)k}{2l+1} {\cal N}_{l+1} \eq 0 \, . \quad (l \geq 2)
\ea
Using $x$ as the time variable and rearranging the equations, we obtain their
final form:
\ba
  \Theta'_0 \eq -\frac{k}{\cal H} \Theta_1 - \Phi' \, , \nonumber \\
  \Theta'_1 \eq \frac{k}{3{\cal H}} \Theta_0 - \frac{2k}{3{\cal H}} \Theta_2 +
    \frac{k}{3{\cal H}} \Psi +
    \tau' \left[ \Theta_1 + \frac{1}{3} v_b \right] , \nonumber \\
  \Theta'_l \eq \frac{lk}{(2l+1){\cal H}} \Theta_{l-1} -
    \frac{(l+1)k}{(2l+1){\cal H}} \Theta_{l+1} \nonumber \\
  \eqnl + \,
    \tau' \left[ \Theta_l - \frac{1}{10} \Pi \, \delta_{l,2} \right] ,
    \qquad l \geq 2 \, , \nonumber \\
  \Theta'_{P0} \eq -\frac{k}{\cal H} \Theta^P_1 +
    \tau' \left[ \Theta^P_0 - \frac{1}{2} \Pi \right] , \nonumber \\
  \Theta'_{Pl} \eq \frac{lk}{(2l+1){\cal H}} \Theta^P_{l-1} -
    \frac{(l+1)k}{(2l+1){\cal H}} \Theta^P_{l+1} \nonumber \\
  \eqnl + \,
    \tau' \left[ \Theta^P_l - \frac{1}{10} \Pi \, \delta_{l,2} \right] ,
    \qquad l \geq 1 \, , \nonumber \\
  {\cal N}'_0 \eq -\frac{k}{{\cal H}} {\cal N}_1 - \Phi' \, , \nonumber \\
  {\cal N}'_1 \eq \frac{k}{3{\cal H}} {\cal N}_0 -
    \frac{2k}{3{\cal H}} {\cal N}_2 + \frac{k}{3{\cal H}} \Psi \, ,
    \nonumber \\
  {\cal N}'_l \eq \frac{lk}{(2l+1){\cal H}} {\cal N}_{l-1} -
    \frac{(l+1)k}{(2l+1){\cal H}} {\cal N}_{l+1} \, , \qquad l \geq 2 \, ,
    \nonumber \\
  \delta' \eq \frac{k}{\cal H} v - 3\Phi' \, , \nonumber \\
  v' \eq -v - \frac{k}{\cal H} \Psi \, , \nonumber \\
  \delta'_b \eq \frac{k}{\cal H} v_b - 3\Phi' \, , \nonumber \\
  v'_b \eq -v_b - \frac{k}{\cal H} \Psi +
    \tau' R \left( 3\Theta_1 + v_b \right) , \nonumber \\
  \Phi' \eq \Psi - \frac{k^2}{3{\cal H}^2} \Phi +
    \frac{H_0^2}{2{\cal H}^2} \left[ \addsmsp{
      \Omega_m a^{-1} \delta + \Omega_b a^{-1} \delta_b
    } \right. \nonumber \\
  && \hspace{29mm} \left. \addsmsp{
      + \, 4\Omega_r a^{-2} \Theta_0 + 4\Omega_\nu a^{-2} {\cal N}_0
    } \right] \! , \nonumber \\
  \Psi \eq -\Phi - \frac{12 H_0^2}{k^2 a^2}
    \left[ \addsmsp{ \Omega_r \Theta_2 + \Omega_\nu {\cal N}_2 } \right] .
  \label{eq:diffeqs}
\ea
The expression for $\Psi$ is just an algebraic equation, so this expression
should simply be inserted into all the other equations when needed. Also, the
expression for $\Phi'$ should be calculated first and used in all the other
equations, so that we obtain a system of differential equations suitable for
the Runge-Kutta method. Note that the only dimensional quantities in
(\ref{eq:diffeqs}) are the wavenumber $k$ and the Hubble function ${\cal H}$.
The natural unit for $k$ is therefore $H_0$. For now, we will ignore the
neutrinos, and return to them later in section~\ref{cha:neutrinos}.

\subsection{Initial conditions}

In order to integrate (\ref{eq:diffeqs}) numerically, we need some initial
conditions at the starting time $x_i = \ln a_i$, where we choose $a_i =
10^{-8}$. Here we consider only adiabatic initial conditions, as derived
in~\cite{Dodelson}:
\ba
  \Theta_0 \eq \frac{1}{2} \Phi \, , \nonumber \\
  \delta \eq \delta_b = \frac{3}{2} \Phi \, , \nonumber \\
  \Theta_1 \eq -\frac{k}{6{\cal H}} \Phi \, , \nonumber \\
  v \eq v_b = \frac{k}{2{\cal H}} \Phi \, .
\ea
The initial condition for $\Phi$ acts as a normalization, and can be chosen to
be $\Phi = 1$~\footnote{This does not mean that the perturbation $\Phi$ of the
gravitational field has the value 1 (remember that $\Phi$ is a small quantity),
but rather that all the other perturbation variables are normalized to the
value of $\Phi$ at $x = x_i$. We also ignore the $k$-dependence of $\Phi$ at
this point, and instead put it back "by hand" in (\ref{eq:Cl_def}).}. Note that
we get
\be
  3\Theta_1 + v_b = 0 \, .
\ee
At early times the optical depth $\tau'$ is very large, meaning that to the
lowest order, everything that is multiplied by $\tau'$ in (\ref{eq:diffeqs})
should be zero. This implies $\Theta_l = 0$ for $l \geq 2$, and $\Theta^P_l =
0$ for all $l$. However, when integrating (\ref{eq:diffeqs}) numerically, we
will need the lowest order non-zero expressions for all the multipoles
(including polarization). And the equations seem to be most well-behaved if we
also use these very small, but non-zero expressions as initial conditions. We
therefore derive these expressions here (see also~\cite{Doran, Zald}).

Very early, the quantity $\epsilon \equiv k/({\cal H} \tau')$ is a small
number, and can therefore be used as an expansion parameter for the multipole
hierarchy. As we will see, $\Theta_l \sim \epsilon \Theta_{l-1}$ for $l \geq
2$, $\Theta^P_0 \sim \Theta^P_2 \sim \Theta_2$, $\Theta^P_1 \sim \epsilon
\Theta_2$, and $\Theta^P_l \sim \epsilon \Theta^P_{l-1}$ for $l \geq 3$.
Assuming that this is true~\footnote{This does not lead to any "circular
logic", since we only assume a certain asymptotic behavior, and then derive
explicit expressions proving that the initial assumption was correct.}, and
also that the derivatives of the multipoles are of the same order as the
multipoles themselves, we can compare the order of magnitude of the different
terms in (\ref{eq:diffeqs}). From $\Theta'_{P0}$ and $\Theta'_{P2}$ we get the
equations $\Theta^P_0 - \Pi/2 = \Theta^P_2 - \Pi/10 = 0$, with the result
\be
  \Theta^P_0 = \frac{5}{4} \Theta_2 \, , \quad
  \Theta^P_2 = \frac{1}{4} \Theta_2 \, , \quad \Rightarrow \quad
  \Pi = \frac{5}{2} \Theta_2 \, .
\ee
Using this in the equation for $\Theta'_2$ we then get
\be
  \Theta_2 = -\frac{8k}{15{\cal H}\tau'} \Theta_1 \, ,
  \label{eq:theta2_tightcoupling}
\ee
and from the equation for~$\Theta'_{P1}$
\be
  \Theta^P_1 = -\frac{k}{4{\cal H}\tau'} \Theta_2 \, .
\ee
Finally, the equations for $l \geq 3$ reduce~to
\ba
  \Theta_l \eq -\frac{l}{2l+1} \frac{k}{{\cal H}\tau'} \Theta_{l-1} \, ,
    \nonumber \\
  \Theta^P_l \eq -\frac{l}{2l+1} \frac{k}{{\cal H}\tau'} \Theta^P_{l-1} \, ,
  \qquad l \geq 3 \, .
\ea
(Remember that $\tau'$ is negative when looking at these expressions.)

\subsection{Tight coupling}

The expressions for the higher temperature and polarization multipoles in the
previous section should be used as long as $k/({\cal H}\tau')$ is small
\footnote{We use the tight coupling approximation as long as $|k/({\cal
H}\tau')| < 1/10$ and $|\tau'| > 10$ (see \cite{Doran}), and switch to the full
equations no later than at the start of recombination (see section
\ref{cha:sourcefunction}).}, which we refer to as the \textit{tight coupling
regime}. However, there's also a more serious numerical problem in this regime,
namely the very small value of $3\Theta_1 + v_b$. This quantity is multiplied
by $\tau'$, which is very large, meaning that even a tiny numerical error in
$\Theta_1$ or $v_b$ will result in completely wrong values for $\Theta'_1$ and
$v'_b$, making the system of differential equations numerically unstable. This
problem can be solved by expanding $3\Theta_1 + v_b$ in powers of $1/\tau'$, as
shown in \cite{Ma_Bert, Doran}. Since this is such an important step in order
to integrate the equations, we include the derivation here.

Playing around with different parts of (\ref{eq:diffeqs}), we get
\ba
  \tau' \! \left( 3\Theta_1 + v_b \right) \eq 3\Theta'_1 +
    \frac{k}{\cal H}(-\Theta_0 + 2\Theta_2) - \frac{k}{\cal H} \Psi \, ,
    \nonumber \\
  (1+R) v'_b \eq -v_b - \frac{k}{\cal H} \Psi \nonumber \\
  && \hspace{-16mm} + \, R \! \left[
    (3\Theta'_1 + v'_b) + \frac{k}{\cal H} (-\Theta_0 + 2\Theta_2) -
    \frac{k}{\cal H} \Psi
  \right] \! , \hspace{6mm} \label{eq:vbderiv} \\
  \Rightarrow \; 3\Theta_1 + v_b \eq \frac{1}{\tau'} \! \left[
    (3\Theta'_1 + v'_b) - v'_b \phantom{\frac{k}{\cal H}}
  \right. \nonumber \\
  && \hspace{4.8mm} \left.
    + \, \frac{k}{\cal H}(-\Theta_0 + 2\Theta_2) -
    \frac{k}{\cal H} \Psi
  \right] \nonumber \\
  \eq \frac{1}{(1+R)\tau'} \! \left[
    (3\Theta'_1 + v'_b) + v_b \phantom{\frac{k}{\cal H}}
  \right. \nonumber \\
  && \hspace{15.7mm} \left.
    + \, \frac{k}{\cal H} (-\Theta_0 + 2\Theta_2)
  \right] \! .
\ea
Taking the derivative of the last equation, using $R' = -R$ and substituting
various expressions for $v_b$ and $v'_b$ so that only the combination
$(3\Theta_1 + v_b)$ and its derivative appear, we get
\ba
  && \hspace{-10mm} \left[ (1+R)\tau' - 1 \right] (3\Theta'_1 + v'_b)
    \nonumber \\
  \eq - \, (1-R)\tau' (3\Theta_1+v_b) - (1+R)\tau'' (3\Theta_1+v_b)
    \nonumber \\
  \eqnl + \, 3\Theta''_1 + v''_b - \frac{k}{\cal H} \Psi +
    \left( 1 - \frac{{\cal H}'}{\cal H} \right) \!
      \frac{k}{\cal H} (-\Theta_0 + 2\Theta_2) \hspace{-10mm} \nonumber \\
  \eqnl + \, \frac{k}{\cal H} (-\Theta'_0 + 2\Theta'_2) \, .
\ea
Until now, everything has been exact. However, during tight coupling it is a
valid approximation to set $3\ddot{\Theta}_1 + \ddot{v}_b$ equal to zero
\cite{Doran}. With $x$ as the variable, this condition turns into
\be
  3\Theta''_1 + v''_b \simeq -\frac{{\cal H}'}{\cal H} (3\Theta'_1 + v'_b) \, .
\ee
This gives us the final expression for $3\Theta'_1 + v'_b$:
\ba
  && \hspace{-10mm} \left[ (1+R)\tau' + \frac{{\cal H}'}{\cal H} - 1 \right]
    (3\Theta'_1 + v'_b) \label{eq:slip} \\
  \eq -\left[ \addsmsp{ (1-R)\tau' + (1+R)\tau'' } \right] (3\Theta_1 + v_b) -
    \frac{k}{\cal H} \Psi \nonumber \\
  \eqnl + \left( 1 - \frac{{\cal H}'}{\cal H} \right) \!
      \frac{k}{\cal H} (-\Theta_0 + 2\Theta_2) +
    \frac{k}{\cal H} (-\Theta'_0 + 2\Theta'_2) \, . \!\! \nonumber
\ea
This expression is then used in (\ref{eq:vbderiv}) to calculate $v'_b$, and
finally $\Theta'_1$ is obtained from
\be
  \Theta'_1 = \frac{1}{3}
    \left[ \addsmsp{ (3\Theta'_1 + v'_b) - v'_b } \right] .
\ee
Note that at early times $\tau' \sim 1/a$, meaning that $\tau'' \simeq -\tau'$.
Therefore, to the leading order, $3\Theta'_1 + v'_b \simeq 2(3\Theta_1 + v_b)
\;\Rightarrow\; 3\Theta_1 + v_b \sim a^2$.

There is one last technical difficulty in this derivation: From (\ref{eq:slip})
we see that $\Theta'_2$ is needed to calculate $3\Theta'_1 + v'_b$ and then
$\Theta'_1$. But from (\ref{eq:theta2_tightcoupling}) we see that $\Theta'_1$
is also needed to calculate $\Theta'_2$. Of course, during tight coupling
$\Theta_2$ is much smaller than $\Theta_0$, so it is probably a good
approximation to simply set $\Theta'_2 = 0$ in (\ref{eq:slip}). Alternatively,
one could use $\Theta'_2 = 0$ as the starting point of a short recurrence
relation where $\Theta'_1$ and $\Theta'_2$ are calculated with growing
precision.

\section{CMB anisotropy spectrum}
\label{cha:cmbspectrum}

When we look at the CMB map today, we are basically observing the values of the
temperature multipoles $\Theta_l(k)$ today. In principle, these can be found by
integrating the system (\ref{eq:diffeqs}) of differential equations from $x_i$
to $x = 0$. However, there are two problems that make this approach very
inefficient. First, we must explicitly include all the multipoles up to the
highest $l$ we're interested in, typically $l \sim 1200$, making the system of
equations extremely large. Secondly, we must integrate the equations for a very
large number of values for $k$, typically several thousand, in order to get an
accurate result. This means that even with todays fast computers, calculating
the CMB anisotropy spectrum would still take many hours. Fortunately, the
calculation time can be reduced by several orders of magnitude by using the
line-of-sight integration method, first developed by Seljak and
Zaldarriaga~\cite{Seljak}.

\subsection{Line-of-sight integration}
\label{cha:lineofsightint}

The basic idea behind the line-of-sight integration method is that instead of
first expanding (\ref{eq:diffeqs_start}) in multipoles and then integrating the
equations, we start by formally integrating the equation for $\dot{\Theta}$ in
(\ref{eq:diffeqs_start}) and do the multipole expansion at the end. As shown in
\cite{Dodelson}, we first get
\ba
  \Theta(k,\mu,\eta_0) \eq \!\! \int_0^{\eta_0} \!\!\left\{ \!
    -\dot{\Phi} - ik\mu\Psi - \dot{\tau} \!\left[
      \Theta_0 + i\mu v_b - \frac{1}{2} {\cal P}_2 \Pi
    \right]\!
  \right\} \hspace{-1mm} \nonumber \\
  && \hspace{3.5mm} \times \, e^{ik\mu(\eta-\eta_0)-\tau} d\eta \, .
  \label{eq:theta0int}
\ea
Because of the exponential, we can replace $ik\mu$ by $d/d\eta$. Using partial
integrations, expanding (\ref{eq:theta0int}) in multipoles, and using the
expression
\be
  \frac{i^l}{2} \int_{-1}^1 {\cal P}_l(\mu) e^{ik\mu(\eta-\eta_0)} d\mu =
    j_l[k(\eta_0-\eta)]
\ee
for the spherical Bessel functions $j_l$, we then get the following expression
for the multipoles today:
\be
  \Theta_l(k,\eta_0) = \int_0^{\eta_0} S(k,\eta) j_l[k(\eta_0-\eta)] d\eta \, .
\ee
The function $S(k,\eta)$ is called the \textit{source function},
\ba
  S(k,\eta) \eq g \left[ \Theta_0 + \Psi + \frac{1}{4} \Pi \right] +
    e^{-\tau} \left[ \dot{\Psi} - \dot{\Phi} \right] \nonumber \\
  \eqnl - \, \frac{1}{k} \frac{d}{d\eta} (g v_b) +
    \frac{3}{4k^2} \frac{d^2}{d\eta^2} (g \Pi) \, .
\ea
With $x$ as the variable, this turns into
\ba
  \Theta_l(k,x=0) \eq \int_{-\infty}^0 \frac{S(k,x)}{{\cal H}(x)}
    j_l[k(\eta_0 - \eta(x))] dx \nonumber \\
  \eqequiv \int_{-\infty}^0
    \tilde{S}(k,x) j_l[k(\eta_0 - \eta(x))] dx \, , \hspace{5mm}
  \label{eq:Thetalint}
\ea
\ba
  \tilde{S}(k,x) \eq
    \tilde{g} \left[ \Theta_0 + \Psi + \frac{1}{4} \Pi \right] +
    e^{-\tau} \left[ \addsmsp{ \Psi' - \Phi' } \right] \nonumber \\
  \eqnl - \, \frac{1}{k} \frac{d}{dx} \left( {\cal H} \tilde{g} v_b \right)
    + \frac{3}{4k^2} \frac{d}{dx} \left[
      {\cal H} \frac{d}{dx} \left( {\cal H} \tilde{g} \Pi \right)
    \right] \! . \hspace{5mm}
  \label{eq:sourcefunction}
\ea
The last term in the source function~is
\ba
  \frac{d}{dx} \left[
    {\cal H} \frac{d}{dx} \left( {\cal H} \tilde{g} \Pi \right)
  \right] \eq
  \frac{d({\cal H}{\cal H}')}{dx} \tilde{g} \Pi +
  3{\cal H}{\cal H}' \left( \tilde{g}'\Pi + \tilde{g}\Pi' \right) \nonumber \\
  \eqnl + \,
  {\cal H}^2 \left( \tilde{g}''\Pi + 2\tilde{g}'\Pi' + \tilde{g}\Pi'' \right) .
\ea
We therefore need the double derivative of $\Pi$ to calculate the source
function. Taking the derivative of the appropriate terms in (\ref{eq:diffeqs}),
we get
\ba
  \Pi'' \! \eq \Theta''_2 + \Theta''_{P2} + \Theta''_{P0} \\
  \eq \frac{d}{dx} \left\{
    \frac{2k}{5{\cal H}} \Theta_1 - \frac{3k}{5{\cal H}} \left(
      \Theta_3 + \Theta^P_1 + \Theta^P_3
    \right) + \frac{3}{10} \tau' \Pi
  \right\} \nonumber \\
  \eq
    \frac{2k}{5{\cal H}}
      \left[ -\frac{{\cal H}'}{\cal H} \Theta_1 + \Theta'_1 \right] +
    \frac{3}{10} \left[ \addsmsp{ \tau'' \Pi + \tau' \Pi' } \right]
      \nonumber \\
  \eqnl - \frac{3k}{5{\cal H}} \!\left[
      -\frac{{\cal H}'}{\cal H} \!
        \left( \Theta_3 \!+\! \Theta^P_1 \!+\! \Theta^P_3 \right) +
      \left( \Theta'_3 \!+\! \Theta'_{P1} \!+\! \Theta'_{P3} \right)
    \right] \! . \nonumber
\ea
Since we need the derivatives of several perturbation variables
$(\Theta'_{1-3}, \Theta'_{P0-3}, \Phi' \;\mathrm{and}\; v'_b)$ in order to
calculate the source function, it can be worthwhile to save these derivatives
along with the variables themselves while integrating the differential
equations, as these derivatives must then be calculated anyway. We are thus
avoiding the need for methods of numerical derivation.

\subsection{Calculating $C_l$}

The observed CMB anisotropy power spectrum today is basically given by $C_l
\sim \Theta^2_l(\vec{x})$ at the point $\vec{x} = 0$, i.e. by the Fourier
transform of $\Theta^2_l(k)$. In addition, since we have so far ignored the
scale-dependence of the initial perturbations, we must also include the
primordial power spectrum $P(k)$. Up to an overall normalization, which we
ignore for now, the CMB power spectrum is therefore given~by
\be
  C_l = \int \frac{d^3 k}{(2\pi)^3} P(k) \Theta^2_l(k) \, .
  \label{eq:Cl_def}
\ee
With a Harrison-Zel'dovich spectrum predicted by inflation, the primordial
power spectrum is
\be
  \frac{k^3}{2\pi^2} P(k) = \left( \frac{k}{H_0} \right)^{n-1} ,
\ee
where $n$ is the spectral index, expected to be close (but not exactly equal)
to 1 from inflation. This gives
\be
  C_l = \int_0^\infty \left( \frac{k}{H_0} \right)^{n-1}
    \Theta^2_l(k) \frac{dk}{k} \, .
  \label{eq:Cl_HZ}
\ee
We will return to normalization in section~\ref{cha:normalization}.

\section{Programming techniques}
\label{cha:programming}

We have already mentioned some of the programming tricks required to be able to
solve all the equations numerically, including integrating the Peebles equation
of recombination, and using the tight coupling approximation. Here we go
through all the other important techniques needed to get both well-behaved and
accurate numerical solutions, and also to get as fast and efficient code as
possible.

\subsection{Diffusion damping -- Boltzmann hierarchy cutoff}
\label{cha:diffusiondamping}

The most obvious thing that has to be done in order to integrate
(\ref{eq:diffeqs}), is to stop the hierarchy of temperature and polarization
multipoles at som maximum $l_{\max}$. If we choose $l_{\max}$ large enough, and
are careful when selecting the cutoff method, there's no need to manually
introduce a damping scale $\sim e^{-k^2/k_D^2}$ like the one used in
\cite{Dodelson}. The easiest cutoff method one can think of is to simply set
$\Theta_{l_{\max}+1} = \Theta^P_{l_{\max}+1} = 0$. However, this method is poor
since power is then transferred from $l_{\max}$ down to $l=0$ and back again on
a timescale $\eta \sim l_{\max}/k$, because of the way the multipoles couple to
each other. A very high value of $l_{\max}$ would therefore be needed to get an
acceptable result, invalidating the whole purpose of the line-of-sight
integration method.

Instead, as discussed in \cite{Ma_Bert}, we look at the time dependence of
$\Theta_l(k,\eta)$ and $\Theta^P_l(k,\eta)$ for large $l$, which is
approximately given~by
\be
  \Theta_l(k,\eta) \, , \, \Theta^P_l(k,\eta) \;\sim\; j_l(k\eta) \, .
\ee
Now remember the recurrence relation for spherical Bessel functions
\be
  j_{l+1}(x) = \frac{2l+1}{x} j_l(x) - j_{l-1}(x) \, .
\ee
It therefore seems plausible to set
\ba
  \Theta_{l+1}(k,\eta) \eqsimeq
    \frac{2l+1}{k\eta} \Theta_l(k,\eta) - \Theta_{l-1}(k,\eta) \, , \\
  \Theta^P_{l+1}(k,\eta) \eqsimeq
    \frac{2l+1}{k\eta} \Theta^P_l(k,\eta) - \Theta^P_{l-1}(k,\eta) \, ,
  \quad l = l_{\max} \, . \nonumber
\ea
Using this approximation in (\ref{eq:diffeqs}) leads~to
\ba
  \Theta'_l \eqsimeq \frac{k}{\cal H} \Theta_{l-1} -
    \frac{l+1}{{\cal H} \eta(x)} \Theta_l + \tau' \Theta_l \, , \\
  \Theta'_{Pl} \eqsimeq \frac{k}{\cal H} \Theta^P_{l-1} -
    \frac{l+1}{{\cal H} \eta(x)} \Theta^P_l + \tau' \Theta^P_l \, ,
  \quad l = l_{\max} \, . \nonumber
\ea
With this cutoff method, even the low value $l_{\max} = 6$ gives a good
agreement with CMBFAST. Of course, this cutoff method is only needed after
tight coupling ends, since during tight coupling all higher multipoles are
expressed directly in terms of the lower ones.

\subsection{Calculating the source function}
\label{cha:sourcefunction}

The source function $\tilde{S}(k,x)$ in (\ref{eq:sourcefunction}) is a smooth
and slowly varying function of both $k$ and $x$, except at the last scattering
surface where it has a sharp peak in $x$. It is therefore sufficient to
integrate the system of differential equations (\ref{eq:diffeqs}) and calculate
$\tilde{S}$ for a rather small number of $k$'s. For each $k$ the result is
stored on a $x$-grid, which has a high resolution during recombination, and a
much lower resolution after recombination \footnote{Here we use the simple
definition that recombination "starts" when $\tilde{g}(x)$ reaches $10^{-20}$
of its maximum value, and "ends" when it is reduced to $0.01$ of the maximum.
For the default model, this gives $z_{\mathrm{start}} = 1630.4$ and
$z_{\mathrm{end}} = 614.2$. Since $\tilde{g}$ falls of exponentially before
recombination we don't need to calculate the source function before this.}.
Choosing 200 points during and 300 points after recombination, evenly
distributed in $x$-space, gives a good agreement with CMBFAST. It is also
sufficient to use 100 different values of $k$ between $k_{\min} = 0.1 H_0$ and
$k_{\max} = 1000 H_0$ (for $l_{\max} = 1200$). A bit of trial and error shows
that we get good results when the $k$'s are distributed quadratically, that is,
$k_i = k_{\min} + (k_{\max} - k_{\min}) (i/100)^2$.

The system of differential equations for each $k$ is integrated using an
adaptive stepsize fifth-order Runge-Kutta method with general Cash-Karp
parameters, as described in \cite{NumRec}. We use a relative error of
$10^{-11}$. The time it takes for the algorithm to finish is roughly
proportional to $k$, with a maximum of a few seconds for $k = 1000 H_0$. The
total time needed to process all 100 $k$-values is therefore about two minutes.
This is by far the most time-consuming part of the calculation.

We will later need the source function also at intermediate values of $k$ and
$x$, that is, we need to make a two-dimensional cubic spline. One way of doing
this is to first take each of the 500 $x$-values and spline across $k$. Then
choose a higher resolution grid of $k$'s, say, 5000 values evenly distributed
between $k_{\min}$ and $k_{\max}$ \footnote{See section \ref{cha:kintegration}
for where the number 5000 comes from.}, and for each $k$ spline across $x$. The
whole splining process still only takes a few seconds to finish. This
two-dimensional spline is also what requires the most memory in the program --
about 120 MB (using 64 bit numbers).

\subsection{Integrating across $x$}
\label{cha:xintegration}

\begin{figure}[!h]
  \begin{center}
    \includegraphics{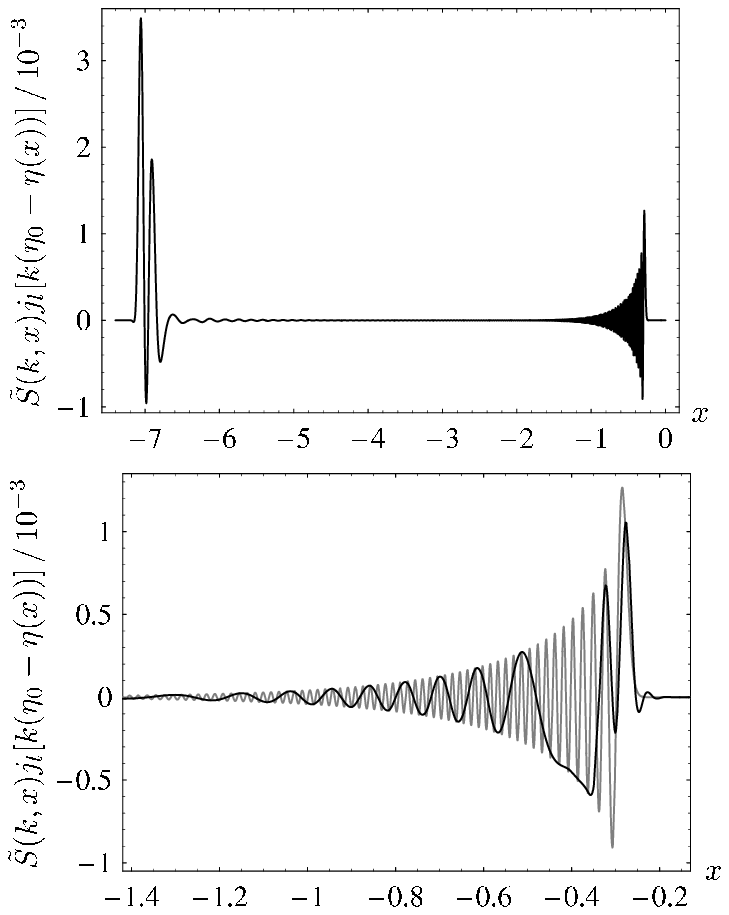}
  \end{center}
  \vspace{-6mm}
  \caption{The integrand in (\ref{eq:Thetalint}) plotted as a function of
  $x$, for $l = 100$ and $k = 340 H_0$. The top figure shows the entire
  integrand from last scattering until today, using the high resolution $x$-
  grid. The bottom figure shows a close-up where the oscillations in the
  Bessel function are important at late times, using the high resolution grid
  (gray line) and low resolution grid (black line). In both cases, we have used
  cubic splining between the actual samples of the integrand.}
  \label{fig:xintegration}
\end{figure}

The source function is smooth in $x$, but the Bessel function $j_l$ in
(\ref{eq:Thetalint}) makes the integrand oscillate for large $k$'s. This may
indicate that we should sample the integrand at more values of $x$ than the 500
points of the grid. We can make a rough estimate of what resolution we should
use: The Bessel function is really just a combination of $\sin$ and $\cos$ with
a period of $2\pi$. This corresponds to an increase in $x$ equal~to
\ba
  && 2\pi \sim k\Delta\eta = k\eta'(x)\Delta x =
    \frac{k}{{\cal H}(x)} \Delta x \, , \nonumber \\
  \Rightarrow && \Delta x \sim \frac{2\pi{\cal H}(x)}{k} \, .
\ea
If we want to have, say, 10 points for each oscillation in the Bessel function,
we must sample the integrand with a resolution
\be
  \Delta x = \frac{2\pi{\cal H}(x)}{10k} \, ,
\ee
i.e. we must use higher resolution at late times (since ${\cal H}(x)$ is a
decreasing function of $x$) and for large $k$'s. We can also make an estimate
of the total number of samples in this grid:
\ba
  && \hspace{-4mm} N \sim \frac{10k}{2\pi} \!\int_{x_{\min}}^0 \!\!
    \frac{dx}{{\cal H}(x)} \sim \frac{10k}{2\pi} \eta_0 \, , \quad
  N(k = 340 H_0) \sim 1800 \, . \nonumber \\
  &&
\ea
Figure \ref{fig:xintegration} shows a part of the integrand for $l = 100$ and
$k = 340 H_0$, compared to the lower resolution grid with 500 points in
$x$-space. Clearly the low resolution grid fails to sample the oscillations in
the Bessel function.

Amazingly, the CMB power spectrum calculated from the low and high resolution
grids are indistinguishable. This is probably because the dominant part of the
$x$-integration comes from recombination, with only a small correction from the
late time oscillations caused by the Bessel function. The source function is
non-zero at late times mostly due to the integrated Sachs-Wolfe effect, which
is most important for low $l$'s ($\sim$ low $k$'s), whereas the oscillations in
the Bessel function dominate for \textit{large} $k$'s, hence the details of the
oscillations are unimportant.

The spherical Bessel functions can be calculated using algorithms described in
\cite{NumRec}. The argument of the function should be chosen from 0 to
$k_{\max} \eta_0 \sim 3400$, and with 10 samples for each oscillation of width
$2\pi$ we need about 5400 samples for each $l$ \footnote{This estimate assumes
$k_{\max} = 1000 H_0$ and $\eta_0 \simeq 3.4 {H_0}\!{}^{-1}$ (the default
model). Since we may need larger values of $k_{\max}$ and get larger $\eta_0$
for other models, it's probably a good idea to use at least twice this maximum
argument, thus sampling the Bessel function at $\sim 10\,000$ points for each
$l$.}. The result is then splined to give a smooth function. The calculation
takes a few seconds for each $l$, but since the Bessel functions are
independent of the cosmological model, they can be calculated once and saved to
disc for fast access later.

Finally, the actual integration across $x$ can be done using either a simple
linear interpolation between the points from the (low or high resolution) grid,
or using a more accurate cubic splining~\footnote{Since we only know the value
of the integrand on a grid of finite resolution, there's no need to resort to
more general methods of numerical integration.}. Also in this case, the
resulting CMB power spectrum is essentially the same, so we therefore choose
the faster linear interpolation.

\subsection{Integrating across $k$ and calculating $C_l$}
\label{cha:kintegration}

The integrand in the final $k$-integration, (\ref{eq:Cl_HZ}), is an oscillating
function of $k$. (This is why we needed the source function for more $k$'s than
the ones where we actually integrated the differential equations.) Since the
dominant contribution to the $x$-integration is from recombination where $\eta
\ll \eta_0$, we have the rough estimate
\be
  \Theta_l(k) \sim j_l(k\eta_0) \int_0^{\eta_0} S(k,\eta) d\eta \sim
    \const \cdot j_l(k\eta_0) \, ,
\ee
since the source function varies much slower with $k$ than the Bessel function.
Thus,
\be
  C_l \sim \int_0^\infty \frac{j_l^2(k\eta_0)}{k} dk \, .
  \label{eq:kint_estimate}
\ee
The Bessel oscillations with period $2\pi$ means that the integrand has
oscillations with period $\Delta k = 2\pi / \eta_0$. In order to sample each
oscillation with 10 points, we must therefore use a grid with resolution
\be
  \Delta k = \frac{2\pi}{10 \eta_0}
\ee
for the $k$-integration. (This leads to the total number of $k$'s
$(10\eta_0/2\pi) (k_{\max}-k_{\min}) \sim 5000$ in
section~\ref{cha:sourcefunction}.)

\begin{figure}[!h]
  \begin{center}
    \includegraphics{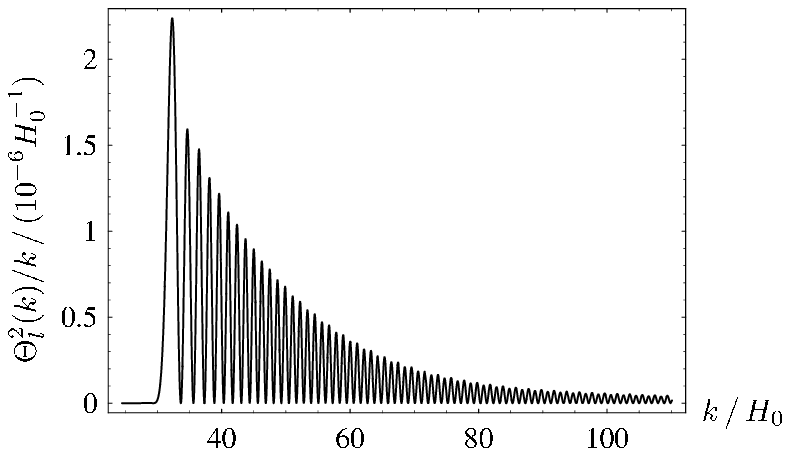}
  \end{center}
  \vspace{-6mm}
  \caption{The integrand in (\ref{eq:Cl_HZ}) plotted as a function of $k$, for
  $l = 100$. The peak is slightly after $k = l/\eta_0 \simeq 29.4 H_0$, and the
  integrand approaches zero very fast for smaller $k$. For larger $k$, however,
  the integrand dies out much more slowly, and even $k_{\max}/H_0 \sim l$ is
  too small for an accurate calculation when $l = 100$. With the adaptive
  algorithm described below, the integration continues all the way to $k =
  454.9 H_0 \sim 15 \, l / \eta_0$.}
  \label{fig:kintegration}
\end{figure}

The computation time can be reduced a bit by observing that the full range $0.1
H_0 \leq k \leq 1000 H_0$ is not needed for all $l$. Instead, from
(\ref{eq:kint_estimate}) we see that the peak of the integrand is around $k
\sim l/\eta_0$. The integrand falls of sharply for smaller $k$, but much more
smoothly for larger $k$, as figure \ref{fig:kintegration} shows. As a first
estimate, we could try the integration range
\be
  \frac{0.9 l}{\eta_0} \leq k \leq \frac{2 l}{\eta_0} \, .
\ee
Using only this interval gives a rather inaccurate result, but we know that
this interval at least contains the peak of the integrand. One possible
algorithm is then to extend the interval in steps of one oscillation of width
$\Delta k = 2\pi/\eta_0$, both to the left and to the right of the initial
interval, and compare the maximum value of the integrand within each step to
the global maximum. We stop when the local maximum has been reduced to less
than, say, $10^{-4}$ of the global. This algorithm gives a CMB power spectrum
identical to using the full interval, but the $k$-integration runs about twice
as fast, since on average we end up using only half of the interval.

Since we only do one $k$-integration for each $l$ (in contrast to several
thousand $x$-integrations), we can use the slower but more accurate cubic
splining instead of linear interpolation for the actual integration. The impact
on the speed of the algorithm from this is negligible.

Finally, we don't need to calculate $C_l$ explicitely for every $l$. Instead,
since the CMB power spectrum is a rather smooth function of $l$, we only
calculate $C_l$ for a few $l$'s and use cubic splining to get a smooth
function~\footnote{Note that in the plots of section \ref{cha:results} we do
not calculate the relative error from the splined $C_l$'s. We use only the
explicitly calculated $C_l$'s and then cubic splining on the relative error
itself. This is because the "artificial" error from the splined $C_l$'s is so
easily removed by simply using more $l$'s, so this does not really indicate any
inaccuracy in the algorithms used.}. For low $l$'s we should use higher
resolution. We choose the points $l = 2$, 3, 4, 6, 8, 10, 12, 15 and 20, and
then every 10th $l$ up to $l = 100$, every 20th up to $l = 200$, every 25th up
to $l = 300$, and finally every 50th $l$ above that. This gives a total of 44
$C_l$-calculations with $l_{\max} = 1200$, and the entire integration (across
both $x$ and $k$) only takes about 20 seconds with precalculated Bessel
functions. Thus, the calculation of the CMB power spectrum is completed in
about two and a half minutes.

\subsection{Normalization}
\label{cha:normalization}

The final point that has to be considered conserns the normalization of the
entire power spectrum. The power spectrum must be properly normalized if we
want to compare it to CMBFAST. If we want to compare just one single model, we
could simply use the height of e.g. the first peak as normalization. However,
since this height depends on cosmological parameters, we must be more careful
if we want to compare several different models within the same figure, like in
figure \ref{fig:powerspectrum1}, \ref{fig:powerspectrum2}
and~\ref{fig:powerspectrum3}.

CMBFAST uses the COBE normalization \cite{Bunn_White}, which basically
normalizes to the observed spectrum from COBE. The idea is to use a least
squares fit of the spectrum for $l \leq 20$~\footnote{More precisely, we use
the points $l = 3$, 4, 6, 8, 12, 15 and 20, since these are also the points
used by CMBFAST.} (since COBE is only accurate on this scale) to a quadratic
function in $x \equiv \log_{10} l$:
\be
  l(l+1) C_l \simeq D_1 \! \left[ 1 + D'(x-1) + D''(x-1)^2/2 \right] .
\ee
From their definition $D'$ and $D''$ are independent of normalization, and
parametrize the shape of the spectrum. The fit to the COBE data is then given
approximately by the formula~\cite{Bunn_White}
\ba
  10^{11} C_{10} \eq 0.64575 + 0.02282 D' + 0.01391 (D')^2 \\
  \eqnl - 0.01819 D'' \!-\! 0.00646 D' D'' \!+\! 0.00103 (D'')^2 , \nonumber
\ea
i.e. the value of $C_{10}$ is fixed by this expression, and the normalization
of the rest of the spectrum then follows by multiplying the calculated $C_l$'s
by the appropriate constant. One should be aware that this normalization may
actually introduce a quite significant uncertainty when comparing the spectrum
to CMBFAST. Part of the reason is that inaccuracies in the calculation of the
low $l$'s get transferred to the entire spectrum by this method. By fine-tuning
the normalization, the relative error in the plots of section \ref{cha:results}
can be reduced by up to a factor of 2. Since in practice what one really want
is to fit the calculated spectrum to observational data, and not to CMBFAST or
some other program, one should probably start using the entire spectrum in the
normalization instead of the COBE normalization. Also note that CMBFAST gives
its output as $l(l+1) C_l / 2\pi$, where the factor $2\pi$ is the commonly used
convention.

\clearpage

\begin{widetext}
\section{Results}
\label{cha:results}

Here we compare our program to CMBFAST for some cosmological models. In figure
\ref{fig:powerspectrum1} we vary the Hubble constant $h$ between $0.66$ and
$0.74$, in figure \ref{fig:powerspectrum2} we vary the baryon density
$\Omega_b$ between $0.042$ and $0.050$, in figure \ref{fig:powerspectrum3} we
vary the CDM density $\Omega_m$ between $0.200$ and $0.248$, and in figure
\ref{fig:powerspectrum4} we vary the spectral index $n$ between $0.95$ and $1$.
In figure \ref{fig:powerspectrum2000} we also plot the power spectrum up to
$l=2000$ for the default model. Following this is the spectrum with helium
included (figure \ref{fig:powerspectrum_helium}, section \ref{cha:helium}), a
simple model of reionization (figure \ref{fig:powerspectrum_reion}, section
\ref{cha:reionization}), massless neutrinos (figure
\ref{fig:powerspectrum_neutrinos}, section \ref{cha:neutrinos}), and finally
the polarization and temperature -- polarization cross correlation power
spectrum (figures \ref{fig:polarization_power} and \ref{fig:cross_power},
section~\ref{cha:polarization}).

\begin{figure}[!h]
  \begin{center}
    \includegraphics{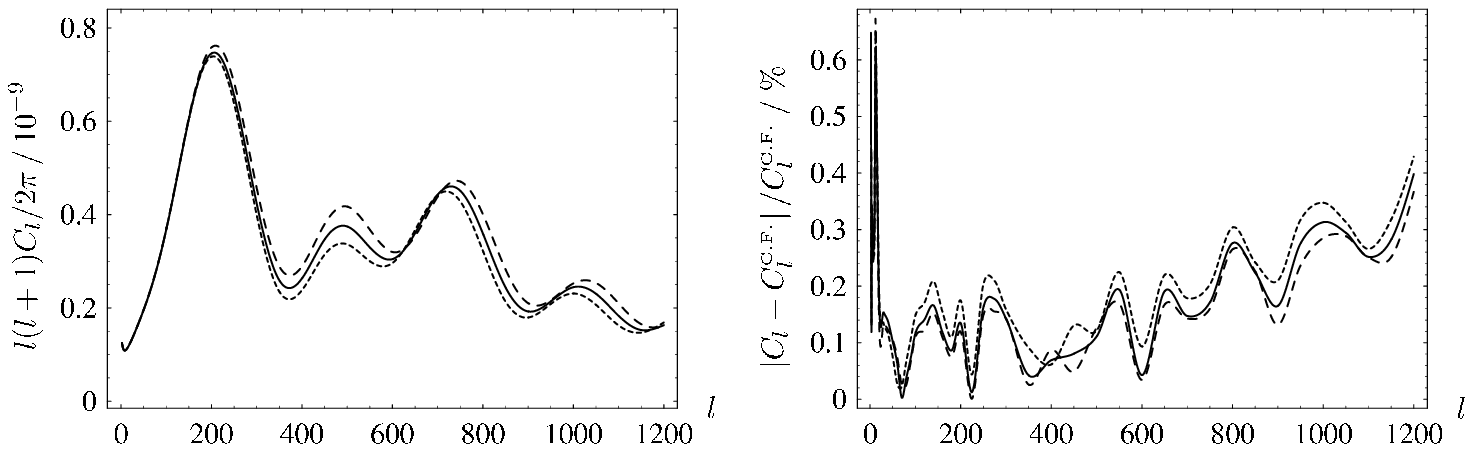}
  \end{center}
  \vspace{-6mm}
  \caption{Our program compared to CMBFAST when varying the Hubble constant:
  the default model $h=0.70$ (solid line), $h=0.66$ (dashed line), and $h=0.74$
  (dotted line). The figure to the left shows our calculated power spectrum,
  and the figure to the right shows the relative error when compared to CMBFAST
  (C.F.), which we see is of order $0.3 \,\%$ or below for most~$l$'s. Because
  the error is so small, there's no point in plotting the spectrum from CMBFAST
  in the figure to the left, since the difference between the curves would be
  smaller than the width of the lines.}
  \label{fig:powerspectrum1}
\end{figure}
\begin{figure}[!h]
  \begin{center}
    \includegraphics{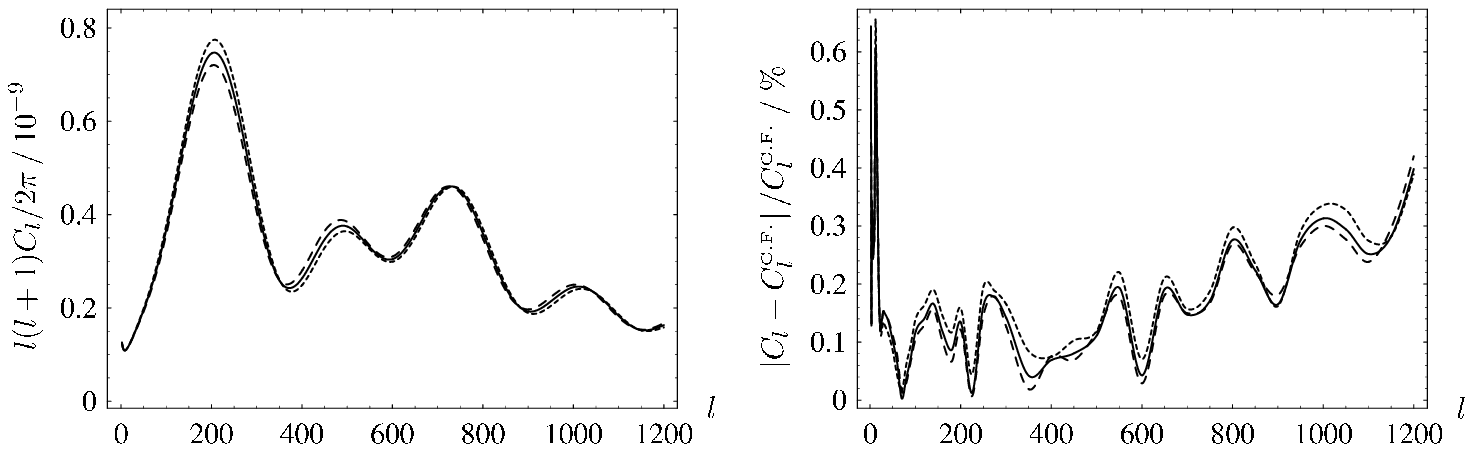}
  \end{center}
  \vspace{-6mm}
  \caption{Our program compared to CMBFAST when varying the baryon density: the
  default model $\Omega_b = 0.046$ (solid line), $\Omega_b = 0.042$ (dashed
  line), and $\Omega_b = 0.050$ (dotted line). The relative error is still
  between $\sim 0.1 \,\%$ and $0.3 \,\%$.}
  \label{fig:powerspectrum2}
\end{figure}
\begin{figure}[!h]
  \begin{center}
    \includegraphics{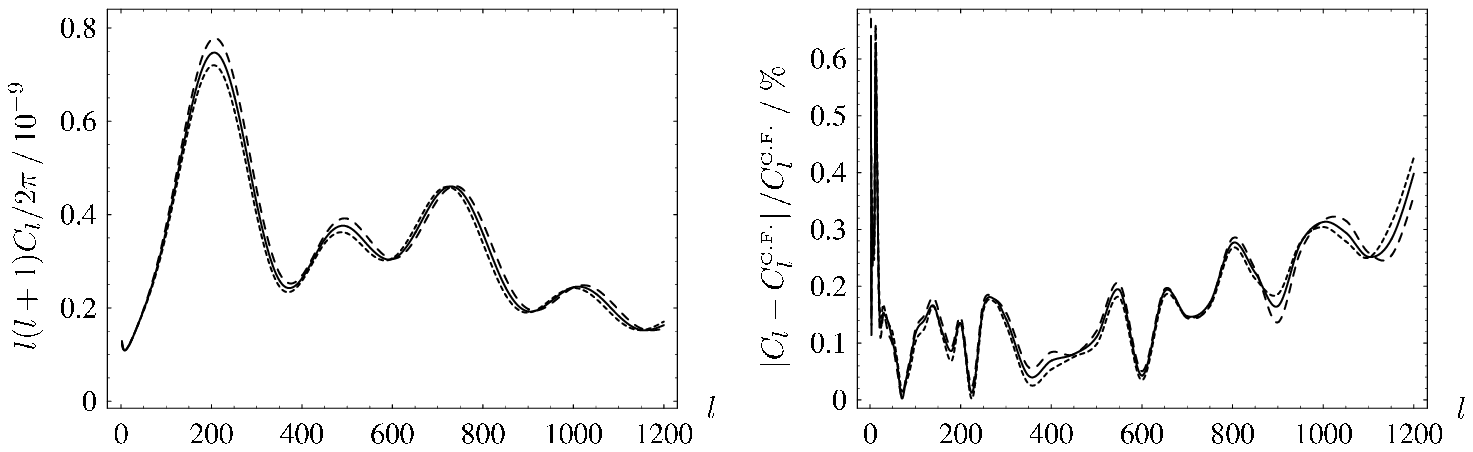}
  \end{center}
  \vspace{-6mm}
  \caption{Our program compared to CMBFAST when varying the CDM density: the
  default model $\Omega_m = 0.224$ (solid line), $\Omega_m = 0.200$ (dashed
  line), and $\Omega_m = 0.248$ (dotted line).}
  \label{fig:powerspectrum3}
\end{figure}

\clearpage

\begin{figure}[!h]
  \begin{center}
    \includegraphics{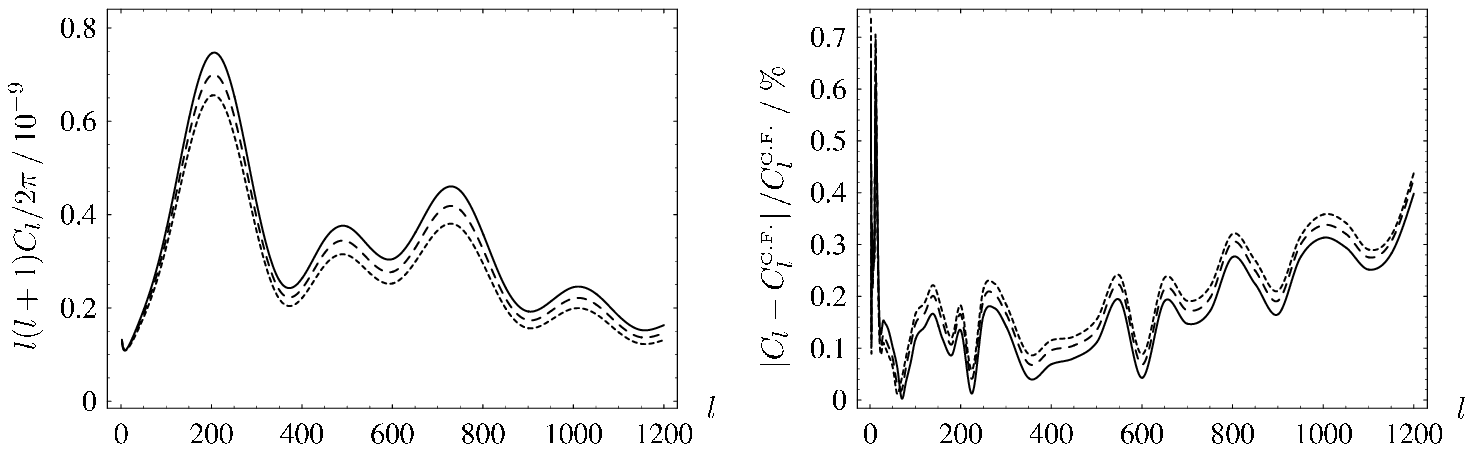}
  \end{center}
  \vspace{-6mm}
  \caption{Our program compared to CMBFAST when varying the spectral index: the
  default model $n = 1$ (solid line), $n = 0.975$ (dashed line), and $n = 0.95$
  (dotted line).}
  \label{fig:powerspectrum4}
\end{figure}
\begin{figure}[!h]
  \begin{center}
    \includegraphics{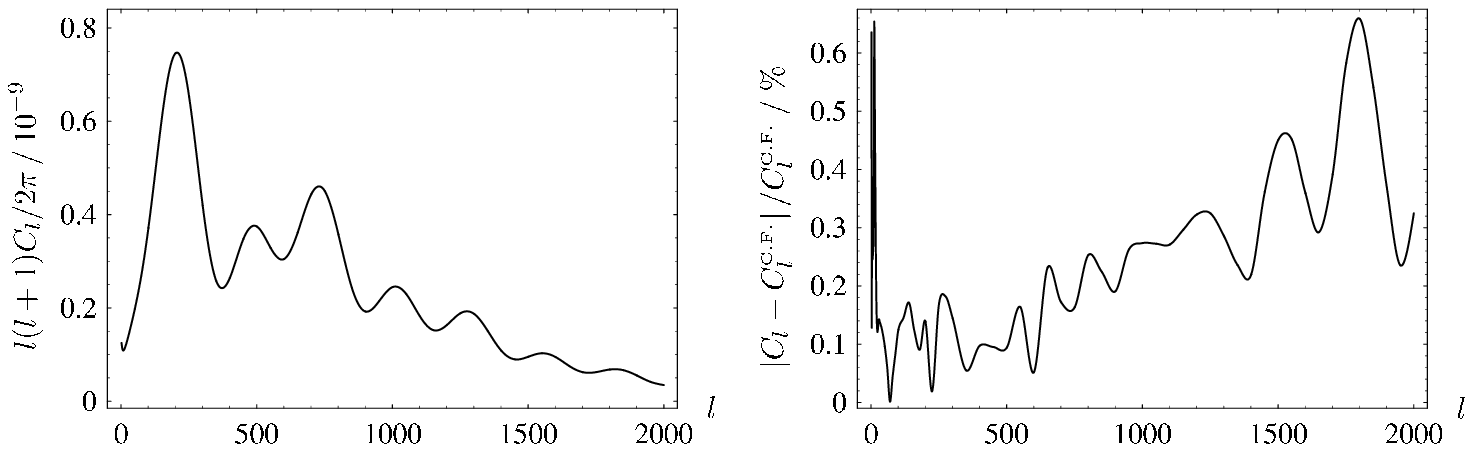}
  \end{center}
  \vspace{-6mm}
  \caption{The power spectrum for the default model up to $l=2000$, compared to
  CMBFAST. The relative error starts to increase systematically beyond
  $l \sim 1000$. In this calculation we have used $k_{\max} = 1500 H_0$, with
  150 values of $k$ chosen initially. We have also used $l_{\max} = 8$ instead
  of $6$ in the Boltzmann hierarchy, which reduced the error for $l \sim 1800$
  from $0.9 \,\%$ down to about~$0.65 \,\%$.}
  \label{fig:powerspectrum2000}
\end{figure}
\begin{figure}[!h]
  \begin{center}
    \includegraphics{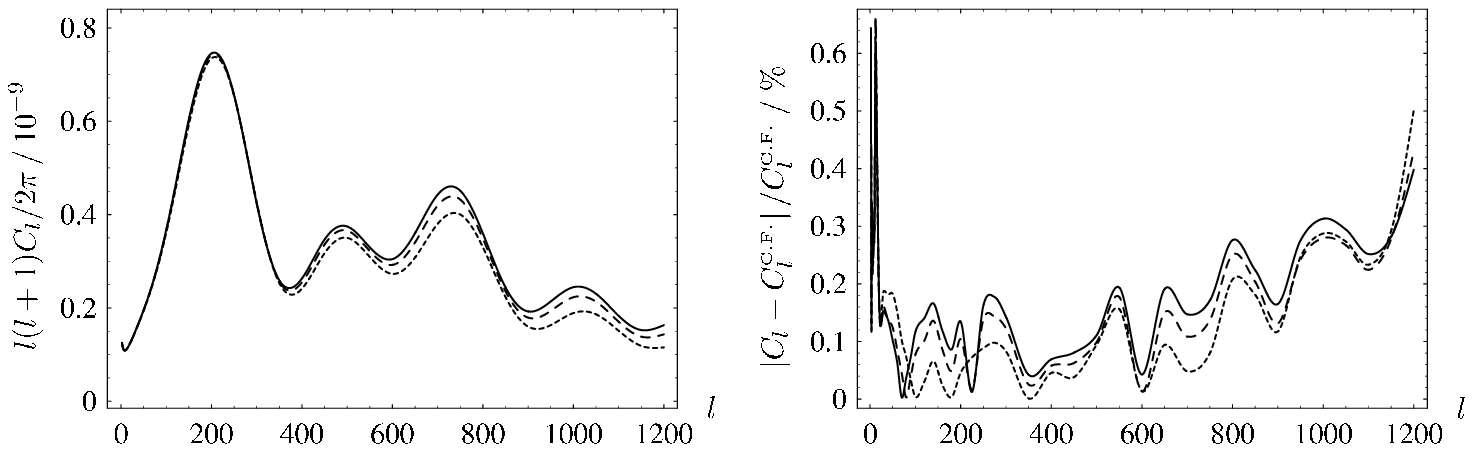}
  \end{center}
  \vspace{-6mm}
  \caption{The power spectrum with helium included (section \ref{cha:helium}),
  compared to CMBFAST: $Y_p = 0$ (solid line), $Y_p = 0.24$ (dashed line), and
  $Y_p = 0.48$ (dotted line). The other cosmological parameters are as in the
  default model.}
  \label{fig:powerspectrum_helium}
\end{figure}

\clearpage

\begin{figure}[!h]
  \begin{center}
    \includegraphics{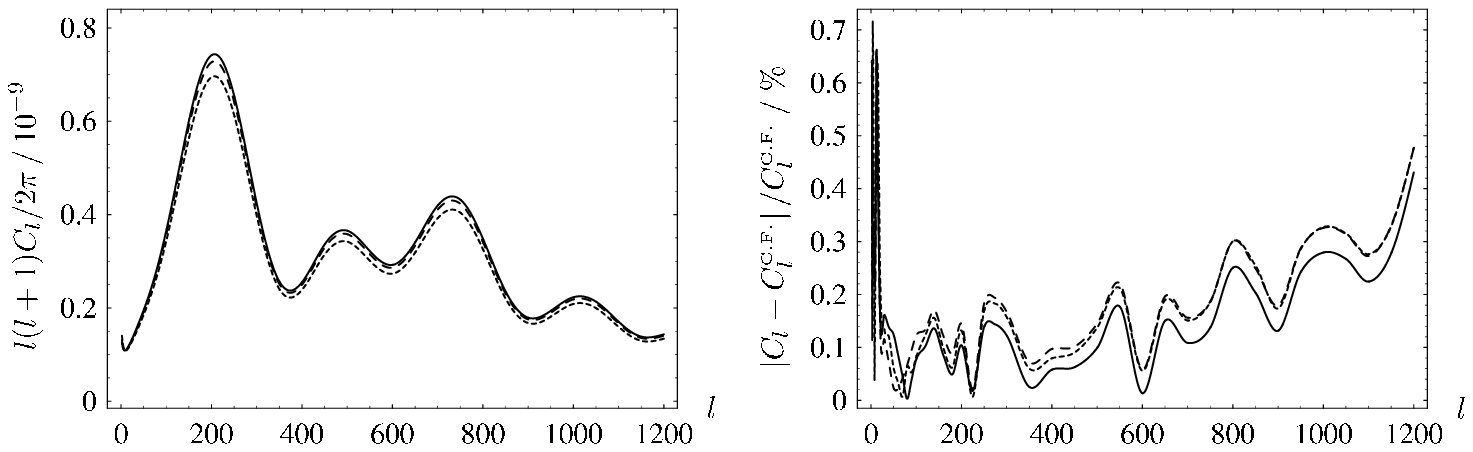}
  \end{center}
  \vspace{-6mm}
  \caption{The power spectrum with reionization (section
  \ref{cha:reionization}): $z_{ri} = 5$ and $\Delta z = 0.075$ (dashed line),
  $z_{ri} = 10$ and $\Delta z = 0.2$ (dotted line), and no reionization
  (solid line). We include helium ($Y_p = 0.24$), otherwise the cosmological
  parameters are as in the default model.}
  \label{fig:powerspectrum_reion}
\end{figure}
\begin{figure}[!h]
  \begin{center}
    \includegraphics{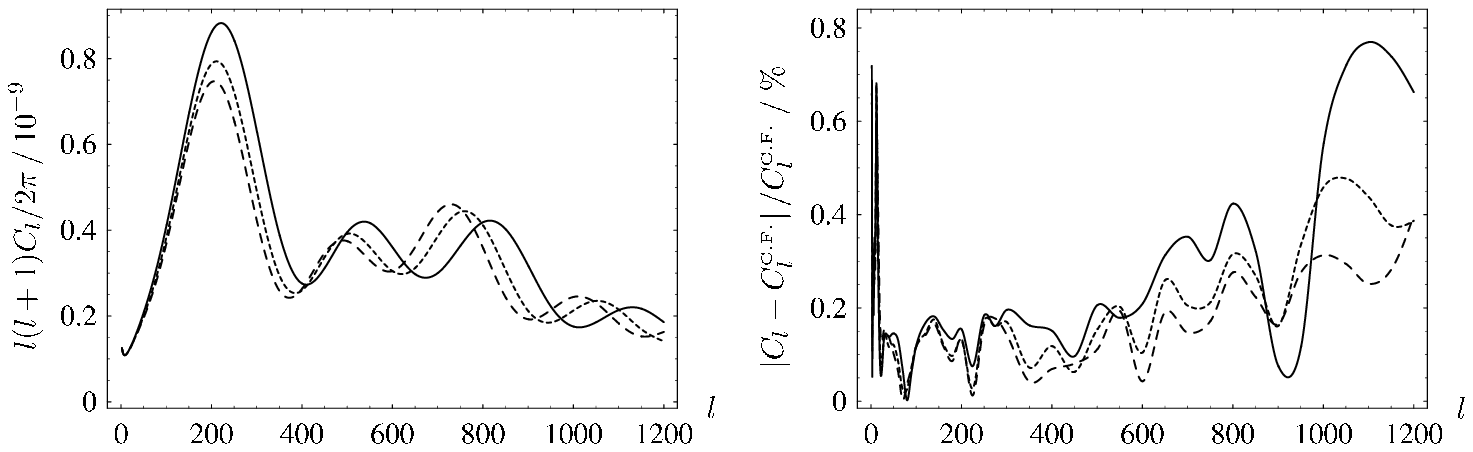}
  \end{center}
  \vspace{-6mm}
  \caption{The power spectrum with massless neutrinos included (section
  \ref{cha:neutrinos}): $N_\nu = 3$ (solid line), $N_\nu = 1$ (dotted line),
  and $N_\nu = 0$, i.e. without neutrinos (dashed line). Apart from the
  neutrinos the default model is used.}
  \label{fig:powerspectrum_neutrinos}
\end{figure}
\begin{figure}[!h]
  \begin{center}
    \includegraphics{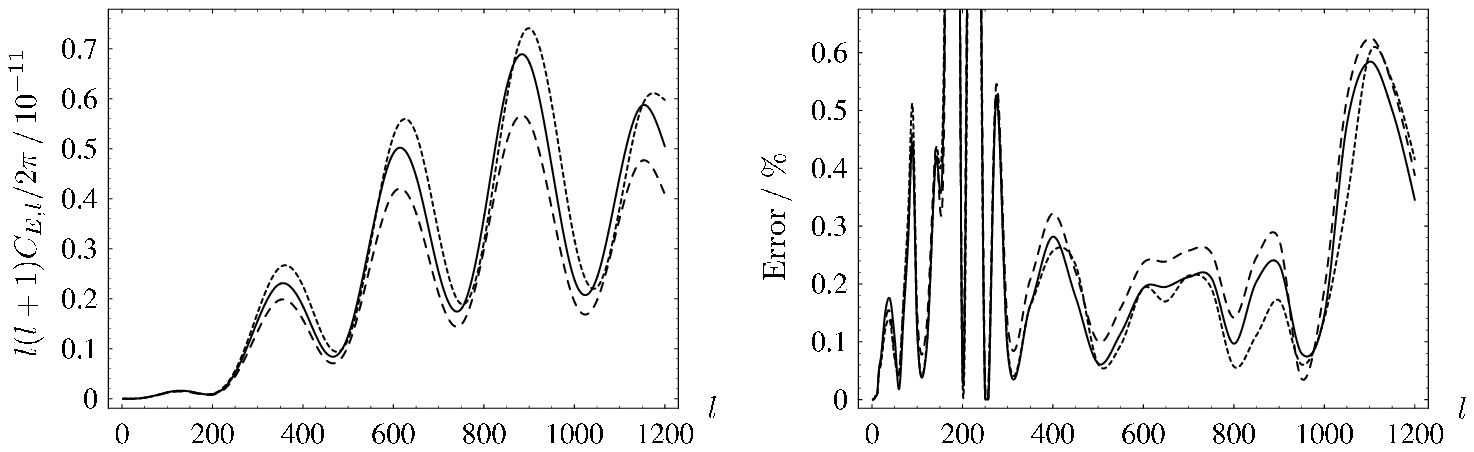}
  \end{center}
  \vspace{-6mm}
  \caption{The E-mode polarization power spectrum (section
  \ref{cha:polarization}) compared to CMBFAST for the default model (solid
  line), $n = 0.95$ (dashed line), and $h = 0.66$ (dotted line). The expression
  for the relative error is slightly modified so it doesn't diverge when
  $C_{E,l} \to 0$: $\mathrm{Error} \equiv |C_{E,l} -
  C_{E,l}^{_{\mathrm{C.F.}}}| \,/\, (C_{E,l}^{_{\mathrm{C.F.}}}\! + \epsilon)$,
  where $l(l+1) \epsilon / 2\pi = 10^{-14}$. The large error of about
  $3 \,\%$ close to $l = 200$ is probably due to a slightly wrong position of
  the minimum. However, the value of $C_{E,l}$ is still very small here, and
  the relative error doesn't really have a significant meaning until $l \gtrsim
  300$.}
  \label{fig:polarization_power}
\end{figure}

\clearpage

\begin{figure}[!h]
  \begin{center}
    \includegraphics{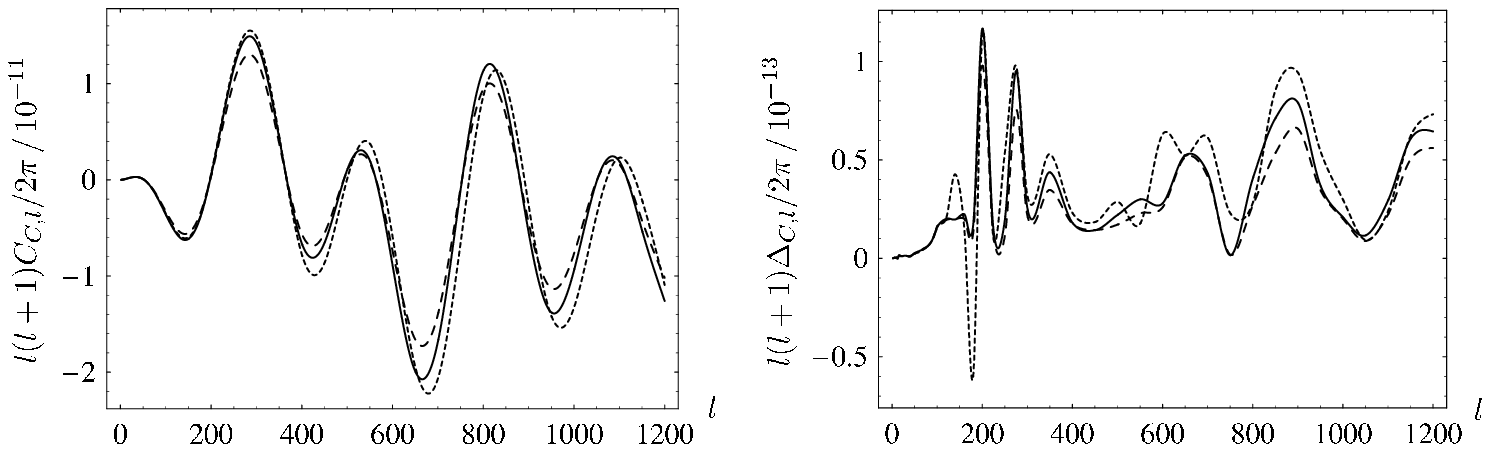}
  \end{center}
  \vspace{-6mm}
  \caption{The temperature -- polarization cross correlation power spectrum
  compared to CMBFAST for the default model (solid line), $n = 0.95$ (dashed
  line), and $h = 0.66$ (dotted line). Since the relative difference is not
  very useful for a function that crosses zero several times, the figure to the
  right shows the absolute difference $\Delta_{C,l} \equiv C_{C,l} -
  C_{C,l}^{_{\mathrm{C.F.}}}$ instead. Note that, with the exception of a
  single point, our program always gives a larger value of $C_{C,l}$ than
  CMBFAST. This can not be due to wrong normalization, however, since $C_{C,l}$
  is both positive and negative.}
  \label{fig:cross_power}
\end{figure}
\end{widetext}

\section{Including more ingredients}
\label{cha:more_ingredients}

\subsection{Helium}
\label{cha:helium}

\begin{figure}[!h]
  \begin{center}
    \includegraphics{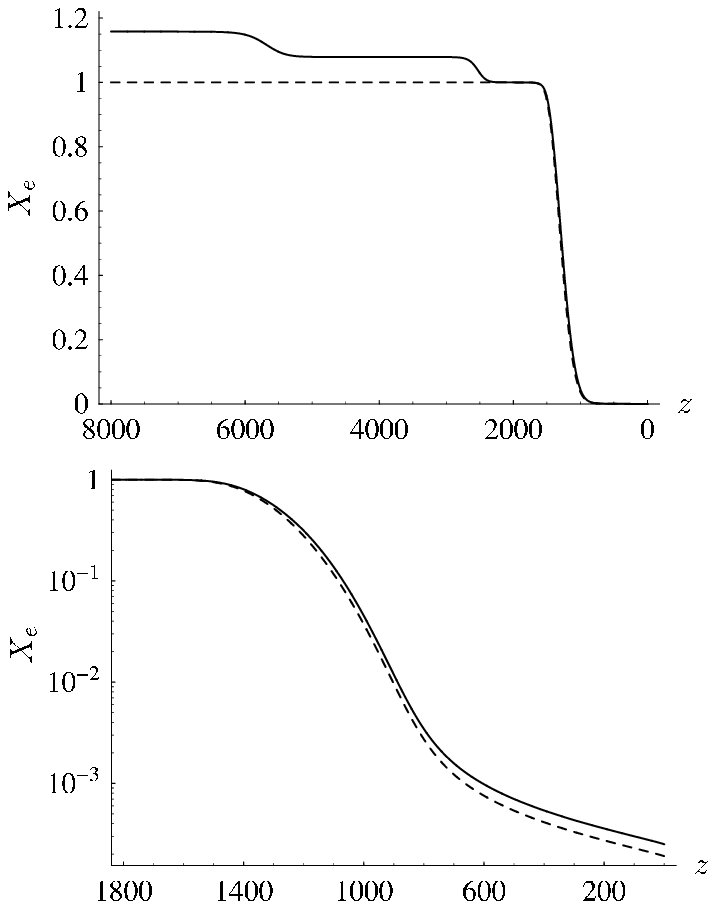}
  \end{center}
  \vspace{-6mm}
  \caption{The electron fraction $X_e$ with helium mass fraction $Y_p = 0.24$
  (solid line) as a function of redshift, compared to $Y_p = 0$ (dashed line).
  The first helium recombination $\he^{++} \to \he^+$ occurs
  around $z \sim 6000$, and the second $\he^+ \to \he$ around
  $z \sim 2500$. For $z \gtrsim 6000$ when helium is doubly ionized $X_e \simeq
  (1-Y_p/2)/(1-Y_p) = 1.1579$, and for $2500 \lesssim z \lesssim 6000$ when
  helium is singly ionized $X_e \simeq (1-3Y_p/4)/(1-Y_p) = 1.0789$. The
  residual electron fraction at late times is larger by a factor $1/(1-Y_p)$
  with helium included.}
  \label{fig:electronfrac_helium}
\end{figure}

The primordial mass fraction of $^4 \he$, $Y_p$, is defined as the ratio
between the total mass of helium and the total baryon mass. We define a
"baryon" as a proton or a neutron with a mass $m_b \simeq m_\hydr$. Each helium
atom thus contains 4 baryons, and has a mass approximately equal to $4
m_\hydr$, which gives~\cite{Trotta_Hansen}
\ba
  Y_p \eqequiv \frac{M_\he}{M_b} \simeq
    \frac{n_\he m_\he}{n_b m_\hydr} \simeq
    \frac{4 n_\he}{n_b} \, , \nonumber \\
  \frac{n_\hydr}{n_b} \eq
    \frac{n_b - 4 n_\he}{n_b} \simeq 1 - Y_p \, .
\ea
Because of the larger ionization energy, helium recombines before hydrogen, so
that during hydrogen recombination all helium is essentially neutral. The main
effect of including helium is therefore that the number of free electrons
during hydrogen recombination is reduced (when keeping $\Omega_b$ fixed). The
recombination history of helium is well described by the Saha equation.
Defining
\be
  x_1 \equiv \frac{n_{\he^+}}{n_\he} \, , \qquad
  x_2 \equiv \frac{n_{\he^{++}}}{n_\he} \, , \qquad
  x_\hydr \equiv \frac{n_{\hydr^+}}{n_\hydr} \, ,
\ee
we now have three Saha equations~\cite{Ma_Bert}
\ba
  n_e \frac{x_1}{1-x_1-x_2} \eq
    2 \left( \frac{m_e T_b}{2\pi} \right)^{3/2} e^{-\chi_0/T_b} \, ,
    \nonumber \\
  n_e \frac{x_2}{x_1} \eq
    4 \left( \frac{m_e T_b}{2\pi} \right)^{3/2} e^{-\chi_1/T_b} \, ,
    \nonumber \\
  n_e \frac{x_\hydr}{1-x_\hydr} \eq
    \left( \frac{m_e T_b}{2\pi} \right)^{3/2} e^{-\epsilon_0/T_b} \, ,
  \label{eq:Saha_helium}
\ea
instead of the one in (\ref{eq:Saha}). They are linked in a non-trivial way,
because the number density of free electrons is now
\ba
  n_e \eq 2 n_{\he^{++}} + n_{\he^+} + n_{\hydr^+}
    \nonumber \\
  \eq \left[ \addsmsp{(2x_2+x_1) Y_p/4 + x_\hydr (1-Y_p)} \right] n_b
    \equiv f_e n_b \, . \hspace{6mm}
  \label{eq:ne_helium}
\ea
The ionization energy of neutral and singly ionized helium~is
\be
  \chi_0 = 24.5874 \;\ev \, , \qquad
  \chi_1 = 4 \,\epsilon_0 = 54.42279 \;\ev \, .
\ee
Eqs. (\ref{eq:Saha_helium}) and (\ref{eq:ne_helium}) are most easily solved by
noting that they will only be used before hydrogen recombination becomes
important, thus $f_e$ is of order 1 the whole time~\footnote{More precisely,
when helium is completely ionized $f_e = 1 - Y_p/2$, whereas after helium
recombination but before hydrogen recombination $f_e \simeq 1 - Y_p$.}. We
therefore use (\ref{eq:Saha_helium}) to express $x_1$, $x_2$ and $x_\hydr$ in
terms of $f_e$, and then use (\ref{eq:ne_helium}) recursively with $f_e = 1$ as
a starting value. The full machine precision of 15 digits is then reached in
less than 10 steps. Finally, the electron fraction $X_e$ defined in
(\ref{eq:electronfrac}) is given~by
\be
  X_e \equiv \frac{n_e}{n_\hydr} = \frac{f_e}{1-Y_p} \, .
\ee
We switch to the more accurate Peebles equation once hydrogen recombination
starts fully ($X_e < 0.99$). At this point all helium is neutral, and the only
difference from section \ref{cha:recombination} is that the hydrogen density is
now smaller than the baryon density. That is, the only changes to eqs.
(\ref{eq:Peebles}) and (\ref{eq:Peebles_def}) are
\ba
  \frac{dX_e}{dx} \eq \frac{C_r(T_b)}{H} \!\left[
    \beta(T_b) (1\!-\!X_e) - (1\!-\!Y_p) n_b \alpha^{(2)}(T_b) X_e^2
  \right] \! , \nonumber \\
  n_{1s} \eq (1-X_e) (1-Y_p) \, n_b \, ,
  \label{eq:Peebles_helium}
\ea
i.e. $n_\hydr$ is replaced by $(1-Y_p) n_b$ everywhere. The resulting solution
$X_e$ is shown in figure \ref{fig:electronfrac_helium} for $Y_p = 0.24$. Note
that $X_e > 1$ before helium recombination.

Once the free electron density and the resulting optical depth have been
calculated, the rest of the CMB calculation proceeds exactly as without helium.
In figure \ref{fig:powerspectrum_helium} we compare our program to CMBFAST for
$Y_p = 0$, $Y_p = 0.24$ and $Y_p = 0.48$ (and the other parameters as in the
default model). The precision with helium included is just as good as without.
One should also note that the other elements (D, $^3\he$, Li etc.) only give
corrections to the CMB power spectrum of order $10^{-5}$~\cite{Hu_Scott}.

\subsection{Reionization}
\label{cha:reionization}

At some time long after recombination, we know that the hydrogen in the
universe became more of less fully ionized again. This was probably the result
of the energetic radiation from the first generation of stars, with enough
energy to ionize hydrogen, but too low energy to ionize helium, which therefore
remained neutral. The detailed mechanism of this process is not fully
understood, but one possible model is to simply assume that at a certain
redshift $z_{ri}$ the free electron fraction $X_e$ instantly jumps to a
constant value, usually $X_e = 1$, and then stays there until today.

\begin{figure}[!h]
  \begin{center}
    \includegraphics{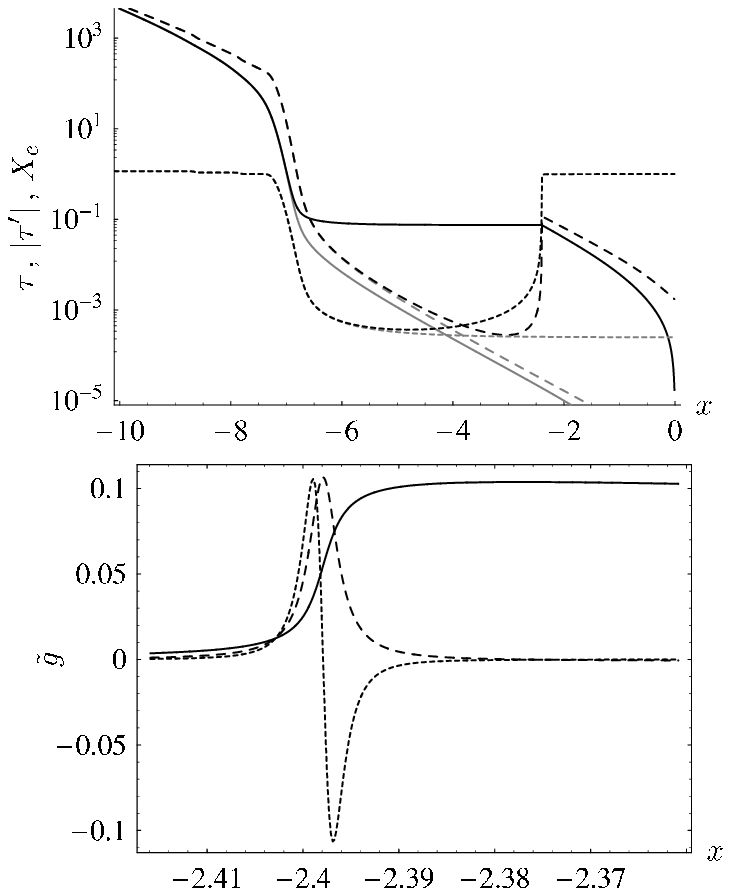}
  \end{center}
  \vspace{-6mm}
  \caption{The top figure shows the optical depth $\tau$ (solid line)
  and $|\tau'|$ (dashed line), and the free electron fraction $X_e$ (dotted
  line), as functions of $x$. The black lines show the semi-instant
  reionization of eq. (\ref{eq:instant_reion}) with $z_{ri} = 10$ and $\Delta z
  = 0.2$, and the gray lines are without reionization, both for the default
  model with helium ($Y_p = 0.24$). The plateau $\tau(z > z_{ri}) \simeq 0.075$
  is the optical depth to the last scattering surface. The bottom figure
  shows the visibility function $\tilde{g} = -\tau' e^{-\tau}$ (solid line),
  its derivative $\tilde{g}'/180$ (dashed line), and its double derivative
  $\tilde{g}''/65\,000$ (dotted line).}
  \label{fig:reionization}
\end{figure}

With instant reionization, both the optical depth $\tau'$ and the visibility
function $\tilde{g}$ experience a jump discontinuity at $z = z_{ri}$, and thus
a delta function in $\tilde{g}'$. This can be a bit tricky to implement
directly in our program, and since it is not very physical either, it is
probably better to use a smooth (but still sharp) transition from the Peebles
result to $X_e = 1$ around $z_{ri}$. We choose the simple formula~\footnote{A
more "natural" choice is probably $f(z) \sim \int e^{-\lambda(z-z_{ri})^2} dz$,
but since this function is not entirely trivial to implement in a program, we
choose the $\arctan$ function instead, which is available directly in most
programming languages.}
\ba
  X_e(z) \eq X^{\mathrm{Peebles}}_e(z) \cdot (1-f) + 1 \cdot f \, ,
    \nonumber \\
  f(z) \eq \frac{1}{\pi}
    \arctan \left[ \frac{10(z_{ri}-z)}{\Delta z} \right] + \frac{1}{2} \, ,
  \label{eq:instant_reion}
\ea
where $\Delta z$ can be interpreted as the width of the reionization period,
typically chosen to be of order $0.2$. Figure \ref{fig:reionization} shows the
free electron fraction, optical depth and visibility function with $z_{ri} =
10$ and $\Delta z = 0.2$.

Because of the sharp peak in $\tilde{g}'$, we must use higher resolution for
the grid of $x$-values near reionization when calculating the source function.
We choose to use an additional 200 points between $z_{ri} + \Delta z$ and
$z_{ri} - \Delta z$, evenly distributed in $x$. The rest of the calculation
then proceeds as without reionization. The resulting CMB power spectrum agrees
very well with CMBFAST, even if CMBFAST uses truly instant reionization. We
should choose $\Delta z$ as small as possible to simulate instant reionization.
Choosing $\Delta z$ too small, however, leads to problems, since the $x$-grid
must then have an even higher resolution, possibly also outside the interval
$\left[ z_{ri} + \Delta z \, , \, z_{ri} - \Delta z \right]$. We get the best
agreement with CMBFAST when $\Delta z = 0.2$ for $z_{ri} = 10$, and $\Delta z =
0.075$ for $z_{ri} = 5$. Figure \ref{fig:powerspectrum_reion} shows the power
spectrum compared to CMBFAST.

\subsection{Massless neutrinos}
\label{cha:neutrinos}

The neutrinos decoupled from the cosmic plasma slightly before the annihilation
of electrons and positrons, when the temperature was of order the electron
mass. The photons were heated by this process, so the neutrino temperature is
therefore lower by a factor~\cite{Dodelson}
\be
  \frac{T_\nu}{T_r} = \left( \frac{4}{11} \right)^{1/3} \simeq 0.714 \, .
\ee
The ratio between the energy densities of neutrinos and photons is
thus~\footnote{Here we have implicitly used the fact that photons have two
polarization degrees of freedom, whereas neutrinos only have one (there are no
right-handed massless neutrinos). However, the neutrino has an antiparticle.
The two species thus have the same number of degrees of freedom, so there are
no additional factors of 2 in~(\ref{eq:rhonu_rhor}).}
\be
  \frac{\rho_\nu}{\rho_r} = \frac{\Omega_\nu}{\Omega_r} =
    N_\nu \cdot \frac{7}{8} \!\left( \!\frac{4}{11}\! \right)^{\!4/3}
    \!\!\! \simeq 0.681 \, ,
    \quad (\mathrm{for}\; N_\nu = 3) \, ,
  \label{eq:rhonu_rhor}
\ee
where $N_\nu$ is the number of neutrino species, and the factor $7/8$ is
because neutrinos are fermions. Actually, since neutrinos are not completely
decoupled when the cosmic plasma is reheated, one should use an effective
number of neutrinos $N_\nu = 3.04$~\cite{Lopez}. With neutrinos as a new
component, the Hubble function (\ref{eq:Hubble}) is of course modified. The
cosmological constant $\Omega_\Lambda$ in (\ref{eq:defaultmodel}) is also
slightly reduced if $\Omega_b$ and $\Omega_m$ are fixed.

The initial conditions for the neutrino monopole and dipole are the same as for
photons:
\ba
  {\cal N}_0 \eq \Theta_0 = \frac{1}{2} \Phi \, , \nonumber \\
  {\cal N}_1 \eq \Theta_1 = -\frac{k}{6{\cal H}} \Phi \, .
\ea
The quadropole is more complicated. Since $\Theta_2 \ll {\cal N}_2$, the
gravitational potentials are initially related by~\cite{Dodelson}
\ba
  \Phi \eq -\Psi \left( 1 + \frac{2 f_\nu}{5} \right) , \\
  f_\nu \eqequiv \frac{\rho_\nu}{\rho_r \!+\! \rho_\nu} =
    \frac{1}{\frac{8}{7 N_\nu\!} \!\left( \frac{11}{4} \right)^{4/3} \!+\! 1}
    \simeq 0.405 \, , \quad (\mathrm{for}\; N_\nu = 3) \, . \nonumber
\ea
From (\ref{eq:diffeqs}) this gives the initial value
\be
  {\cal N}_2 = -\frac{k^2 a^2 \Phi}{12 H_0^2 \Omega_\nu}
    \frac{1}{\frac{5}{2 f_\nu} + 1} \, .
  \label{eq:N2_early}
\ee
Note that ${\cal N}_2 \sim a^2$ at early times. For the higher multipoles we
can assume that ${\cal N}_l \ll {\cal N}_{l-1}$. Using (\ref{eq:N2_early}) and
${\cal H} \simeq H_0 \sqrt{\Omega_r+\Omega_\nu} \, a^{-1}$ in
(\ref{eq:diffeqs}) then gives a differential equation for ${\cal N}_3$ that can
be integrated directly, with the result that ${\cal N}_3 \sim a^3$. Continuing
this way we get ${\cal N}_l \sim a^l$, meaning that ${\cal N}'_l \simeq l {\cal
N}_l$, and therefore
\be
  {\cal N}_l \simeq \frac{k}{(2l+1){\cal H}} {\cal N}_{l-1} \, ,
  \qquad l \geq 3 \, ,
  \label{eq:Nl_early}
\ee
as the initial condition for the higher multipoles. Finally, we use the same
cutoff scheme for the neutrino hierarchy as for the photons,
\be
  {\cal N}'_l \simeq \frac{k}{{\cal H}} {\cal N}_{l-1} -
    \frac{l+1}{{\cal H}\eta(x)} {\cal N}_l \, , \qquad l = l_{\max} \, .
  \label{eq:cutoff_neutrinos}
\ee
We choose $l_{\max} = 10$ for the neutrino multipoles in order to get
sufficient precision for large $l$'s. Note that ${\cal H} \eta \simeq 1$ at
early times, so that (\ref{eq:cutoff_neutrinos}) reduces to ${\cal N}'_l \simeq
l {\cal N}_l$ directly when using (\ref{eq:Nl_early}).

As a curiosity we find that
\be
  \lim_{N_\nu \to 0} {\cal N}_2^{\mathrm{init}} =
    -\frac{k^2 a^2 \Phi}{30 H_0^2 \Omega_r} \, .
\ee
All the equations where neutrinos appear are therefore well-defined in the
limit $N_\nu \to 0$. The hierarchy of neutrino multipoles is still
non-vanishing in this limit, but the neutrinos no longer contribute to the CMB
power spectrum since they decouple from the gravitational potential. The result
is therefore the same as if the neutrino hierarchy had not been included at
all, as expected.

The rest of the CMB calculation is exactly the same as without neutrinos, as
long as one uses the correct expression for $\Psi$ in the source function
(\ref{eq:sourcefunction}). Figure \ref{fig:powerspectrum_neutrinos} shows the
power spectrum for $N_\nu = 3$, $1$ and $0$ compared to CMBFAST. The agreement
is very good below $l \sim 600$, but the error increases somewhat faster for
large $l$'s than without neutrinos.

\subsection{Polarization power spectrum}
\label{cha:polarization}

So far, we have only considered the temperature power spectrum of the CMB.
However, since the radiation is polarized, we also have both a polarization
power spectrum and a cross correlation power spectrum between temperature and
polarization. With the recent release of the three-year results of
WMAP~\cite{WMAP_threeyear}, these spectra will be of great interest since we
now have measurements of the full sky CMB polarization map.

A radiation field in general needs four parameters to be described completely,
called the Stokes parameters. These are the temperature~$T$, linear
polarization $Q$ and $U$ along two different directions, and circular
polarization~$V$. $T$ and $V$ are rotationally invariant and can therefore be
expanded in spherical harmonics~\footnote{Circular polarization can not be
generated through Thomson scattering, so we will ignore $V$ from now on.}. $Q$
and $U$, on the other hand, transform under rotations in the plane
perpendicular to the direction of the photons. It turns out that the linear
combination $Q \pm iU$ transforms in a particularly simple way. Under a
rotation $\psi$ it transforms as $(Q \pm iU)' = e^{\mp 2i\psi} (Q \pm iU)$,
i.e. it has spin~$\pm 2$ and can therefore be expanded in what is called
spin~$\pm 2$ spherical harmonics~\footnote{An alternative method is to
construct a $2 \times 2$ symmetric traceless tensor from $Q$ and $U$, and
expand this in tensor spherical harmonics~\cite{KKS}.} (see \cite{Zald_Seljak}
for more details). It also means that we can define spin zero quantities by
acting on $Q \pm iU$ twice using the spin raising operator $\eth$ or the spin
lowering operator $\overline{\eth}$ (again see \cite{Zald_Seljak} for the
details)
\ba
  \tilde{E}(\mu) \eqequiv -\frac{1}{2} \left[ \addsmsp{
    \overline{\eth} {}^{\:\! 2} (Q+iU) + \eth^2(Q-iU)
  } \right] , \nonumber \\
  \tilde{B}(\mu) \eqequiv \frac{i}{2} \left[ \addsmsp{
    \overline{\eth} {}^{\:\! 2} (Q+iU) - \eth^2(Q-iU)
  } \right] .
\ea
The power spectra for $\tilde{E}$ and $\tilde{B}$ are thus rotationally
invariant, and can be used to describe the polarization of the CMB radiation.

When we are only considering scalar perturbations, we can choose a coordinate
system (for each Fourier mode) where $U = 0$. We then have $Q = \Theta_P$ and
$\overline{\eth} {}^{\:\! 2} Q = \eth^2 Q$ since $\Theta_P$ only depends on the
polar angle. Thus we only get E-mode polarization from scalar perturbations.
(Tensor perturbations generate both E- and B-mode polarization.) We use the
line-of-sight integration method for polarization, similar to the temperature,
and get from~(\ref{eq:diffeqs_start})
\ba
  \Theta_P(k,\mu,\eta_0) \eq -\frac{1}{2} \int_0^{\eta_0} \dot{\tau}
    (1 - {\cal P}_2) \Pi e^{ik\mu(\eta-\eta_0) - \tau} d\eta \nonumber \\
  \eq \frac{3}{4} \int_0^{\eta_0} g \Pi (1-\mu^2)
    e^{ik\mu(\eta-\eta_0)} d\eta \, .
\ea
This gives~\cite{Zald_Seljak}
\ba
  \tilde{E}(k,\mu,\eta_0) \eq -\frac{3}{4} \int_0^{\eta_0} g \Pi
    \partial_\mu^2 \left[ (1-\mu^2)^2 e^{-i\mu x} \right] \!d\eta \nonumber \\
  \eq \frac{3}{4} \int_0^{\eta_0} g \Pi
    (1 + \partial_x^2)^2 (x^2 e^{-i\mu x}) \, d\eta \, ,
\ea
where $x = k(\eta_0 - \eta)$. Expanding in multipoles, we get~\footnote{See
\cite{Zald_Seljak} for the details on the extra factor $\sqrt{(l-2)! /
(l+2)!}$.}
\ba
  \Theta^E_l(k,\eta_0) \eq \frac{3}{4} \sqrt{\frac{(l-2)!}{(l+2)!}}
    \int_0^{\eta_0} g \Pi (1 + \partial_x^2)^2
    \!\left[ x^2 j_l(x) \right]\! d\eta \nonumber \\
  \eq \frac{3}{4} \sqrt{\frac{(l+2)!}{(l-2)!}} \int_0^{\eta_0} g \Pi
    \frac{j_l(x)}{x^2} d\eta \nonumber \\
  \eq \sqrt{\frac{(l+2)!}{(l-2)!}} \int_0^{\eta_0}
    S_E(k,\eta) j_l[k(\eta_0-\eta)] d\eta \, , \nonumber \\
  S_E(k,\eta) \eq \frac{3g\Pi}{4k^2(\eta_0-\eta)^2} \, .
\ea
Here we have used the result $(1 + \partial_x^2)^2 \!\left[ x^2 j_l(x)
\right]\! = (l-1)l(l+1)(l+2) j_l(x) / x^2$, which follows from the differential
equation $j''_l + 2 j'_l / x + \!\left[ 1 - l(l+1)/x^2 \right]\! j_l = 0$
satisfied by the spherical Bessel function. The E-mode polarization power
spectrum and its cross correlation with temperature is then finally given
by~\cite{Zald_Seljak}
\ba
  C_{E,l} \eq \int_0^\infty \left( \frac{k}{H_0} \right)^{n-1}
    \Theta_{El}^2(k) \frac{dk}{k} \, , \nonumber \\
  C_{C,l} \eq \int_0^\infty \left( \frac{k}{H_0} \right)^{n-1}
    \Theta_l(k) \Theta^E_l(k) \frac{dk}{k} \, .
  \label{eq:polpower_E}
\ea
Figure \ref{fig:polarization_power} shows the E-mode polarization and figure
\ref{fig:cross_power} the temperature -- polarization cross correlation
compared to CMBFAST for a few models. The calculation uses exactly the same
algorithms and techniques as for the temperature, only with $l_{\max} = 8$
instead of $6$ in the Boltzmann hierarchy to get acceptable precision for the
polarization multipoles.

\section{Conclusion}
\label{cha:conclusion}

We have here presented all the main steps required in writing a program that
calculates the CMB anisotropy power spectrum. Our focus has been on the
computer-technical side of the problem, by including all the small details that
make the program actually work, something which is often left out in the
literature. We have consentrated on the $\lcdm$ model, where the program
achieves an accuracy comparable to CMBFAST~\footnote{The accuracy of CMBFAST is
of order $0.1 \,\%$~\cite{CMBFAST_accuracy}.} ($\sim \, 0.1 \,\%$ -- $0.4
\,\%$) over a range of cosmological parameters. The program runs in a couple of
minutes on a mid-range personal computer (as of 2006). While certainly not as
good as CMBFAST, this is still acceptable considering that the code hasn't
really been optimized for speed.

The purpose of this work has been to give a running start to those needing to
calculate the CMB power spectrum for some exotic cosmological model where the
standard programs can't be used. With the growing precision of the observed
spectrum, a calculation to within the $1 \,\%$ level is often what
distinguishes the models and makes it possible to rule out some of them. We
hope this work will encourage others by showing that writing a program from
scratch to within this accuracy is not really as difficult or time-consuming as
one may think.

\textbf{Acknowledgement:} I want to thank Tomi Koivisto for very useful
discussions about the various programming techniques involved in writing the
program. This work has been supported by grant no. NFR 153577/432 from the
Research Council of Norway.


\begin{thebibliography}{10}

\bibitem{COBE}             
  G. F. Smoot \textit{et al.}, Astrophys. J. Lett. \textbf{396}, L1 (1992)
\bibitem{WMAP_firstyear}   
  C. L. Bennett \textit{et al.},
  Astrophys. J. Suppl. \textbf{148}, 1 (2003), astro-ph/0302207;
  D. N. Spergel \textit{et al.},
  Astrophys. J. Suppl. \textbf{148}, 175 (2003), astro-ph/0302209
\bibitem{WMAP_threeyear}   
  D. N. Spergel \textit{et al.}, astro-ph/0603449;
  L. Page \textit{et al.}, astro-ph/0603450;
  G. Hinshaw \textit{et al.}, astro-ph/0603451;
  N. Jarosik \textit{et al.}, astro-ph/0603452
\bibitem{Planck}           
  See the Planck satellite webpage at\\
  http://www.rssd.esa.int/index.php?project=planck or
  http://www.esa.int/science/planck
\bibitem{StandardModel}    
   D. Scott, astro-ph/0510731
\bibitem{CMBFAST}          
  The CMBFAST program package is written by U. Seljak and M. Zaldarriaga. See
  its webpage at\\
  http://www.cmbfast.org
\bibitem{CMBFAST_web}      
  A web interface of CMBFAST can be found at\\
  http://lambda.gsfc.nasa.gov/cgi-bin/cmbfast\_form.pl. All spectra from
  CMBFAST in this text have been obtained from this web form.
\bibitem{CMBEASY}          
  M. Doran, JCAP \textbf{0510}, 011 (2005), astro-ph/0302138;
  See also its webpage http://www.cmbeasy.org
\bibitem{CAMB}             
  A. Lewis, A. Challinor and A. Lasenby,
  Astrophys. J. \textbf{538}, 473 (2000), astro-ph/9911177;
  See also its webpage http://camb.info
\bibitem{Delphi}           
  http://www.borland.com/delphi
\bibitem{Dodelson}         
  S. Dodelson, \textit{Modern cosmology}, Academic Press (2003)
\bibitem{Ma_Bert}          
  C.-P. Ma and E. Bertschinger, Astrophys. J. \textbf{455}, 7 (1995),
  astro-ph/9506072
\bibitem{Hu}               
  W. Hu, Annals Phys. \textbf{303}, 203 (2003), astro-ph/0210696
\bibitem{Reijo}            
  R. Keskitalo, Master thesis, University of Helsinki, Helsinki (2005)
\bibitem{Doran}            
  M. Doran, JCAP \textbf{0506}, 011 (2005), astro-ph/0503277
\bibitem{Zald}             
  M. Zaldarriaga and D. D. Harari,
  Phys. Rev. \textbf{D52}, 3276 (1995), astro-ph/9504085
\bibitem{Seljak}           
  U. Seljak and M. Zaldarriaga, Astrophys. J. \textbf{469}, 437 (1996),
  astro-ph/9603033
\bibitem{NumRec}           
  W. H. Press, S. A. Teukolsky, W. T. Vetterling and B. P. Flannery,
  \textit{Numerical Recipes in C}, Cambridge University Press (2002)
\bibitem{Bunn_White}       
  E. F. Bunn and M. White,
  Astrophys. J. \textbf{480}, 6 (1997), astro-ph/9607060
\bibitem{Trotta_Hansen}    
  R. Trotta and S. H. Hansen,
  Phys. Rev. \textbf{D69}, 023509 (2004), astro-ph/0306588
\bibitem{Hu_Scott}         
  W. Hu, D. Scott, N. Sugiyama and M. White,
  Phys. Rev. \textbf{D52}, 5498 (1995), astro-ph/9505043
\bibitem{Lopez}            
  R. E. Lopez, S. Dodelson, A. Heckler and M. S. Turner,
  Phys. Rev. Lett. \textbf{82}, 3952 (1999), astro-ph/9803095
\bibitem{Zald_Seljak}      
  M. Zaldarriaga and U. Seljak,
  Phys. Rev. \textbf{D55}, 1830 (1997), astro-ph/9609170
\bibitem{KKS}              
  M. Kamionkowski, A. Kosowsky and A. Stebbins,
  Phys. Rev. \textbf{D55}, 7368 (1997), astro-ph/9611125
\bibitem{CMBFAST_accuracy} 
  U. Seljak, N. Sugiyama, M. White and M. Zaldarriaga,
  Phys. Rev. \textbf{D68}, 083507 (2003), astro-ph/0306052



\end{thebibliography}
\end{document}